\documentclass[twocolumn]{aastex62}
\usepackage[utf8]{inputenc}
\usepackage{amsmath}
\usepackage{natbib}
\usepackage{graphicx}
\usepackage{float}
\usepackage{hyperref}
\usepackage{soul}
\soulregister\cite7 % 针对\cite命令
\soulregister\citep7 % 针对\citep命令
\soulregister\citet7 % 针对\citet命令
\soulregister\ref7 % 针对\ref命令
\soulregister\pageref7 % 针对\pageref命令

\begin{document}

\title{First SETI Observations with China's Five-hundred-meter Aperture Spherical radio Telescope (FAST)}

\author{Zhi-Song Zhang}
\affiliation{National Astronomical Observatories, Chinese Academy of Sciences, Beijing 100012, China}
\affiliation{University of Chinese Academy of Sciences, Beijing 100049, China}
\affiliation{Space Sciences Laboratory, University of California, Berkeley, Berkeley CA 94720}%https://www.overleaf.com/3532882631gdzbssnckxfx
\affiliation{Department of Astronomy, University of California Berkeley, Berkeley CA 94720, USA}

\author{Dan Werthimer}
\affiliation{Space Sciences Laboratory, University of California, Berkeley, Berkeley CA 94720}
\affiliation{Department of Astronomy, University of California Berkeley, Berkeley CA 94720, USA}

\author{Tong-Jie Zhang}
\affiliation{Department of Astronomy, Beijing Normal University, Beijing 100875, China}

\author{Jeff Cobb}
\affiliation{Space Sciences Laboratory, University of California, Berkeley, Berkeley CA 94720}
\affiliation{Department of Astronomy, University of California Berkeley, Berkeley CA 94720, USA}

\author[0000-0001-8078-9395]{Eric Korpela}
\affiliation{Space Sciences Laboratory, University of California, Berkeley, Berkeley CA 94720}

\author{David Anderson}
\affiliation{Space Sciences Laboratory, University of California, Berkeley, Berkeley CA 94720}

\author{Vishal Gajjar}
\affiliation{Space Sciences Laboratory, University of California, Berkeley, Berkeley CA 94720}
\affiliation{Department of Astronomy, University of California Berkeley, Berkeley CA 94720, USA}

\author{Ryan Lee}
\affiliation{Department of Astronomy, University of California Berkeley, Berkeley CA 94720, USA}
\affiliation{Department of Physics, University of California Berkeley, Berkeley CA 94720, USA} 
\affiliation{Department of Computer Science, University of California Berkeley, Berkeley CA 94720, USA} 

\author{Shi-Yu Li}
\affiliation{Department of Astronomy, Beijing Normal University, Beijing 100875, China}

\author{Xin Pei}
\affiliation{University of Chinese Academy of Sciences, Beijing 100049, China}
\affiliation{Xinjiang Astronomical Observatory, CAS, 150, Science 1-Street, Urumqi, Xinjiang 830011, China}

\author{Xin-Xin Zhang}
\affiliation{National Astronomical Observatories, Chinese Academy of Sciences, Beijing 100012, China}

\author{Shi-Jie Huang}
\affiliation{National Astronomical Observatories, Chinese Academy of Sciences, Beijing 100012, China}

\author{Pei Wang}
\affiliation{National Astronomical Observatories, Chinese Academy of Sciences, Beijing 100012, China}

\author{Yan Zhu}
\affiliation{National Astronomical Observatories, Chinese Academy of Sciences, Beijing 100012, China}

\author{Ran Duan}
\affiliation{National Astronomical Observatories, Chinese Academy of Sciences, Beijing 100012, China}

\author{Hai-Yan Zhang}
\affiliation{National Astronomical Observatories, Chinese Academy of Sciences, Beijing 100012, China}

\author{Cheng-jin Jin}
\affiliation{National Astronomical Observatories, Chinese Academy of Sciences, Beijing 100012, China}

\author{Li-Chun Zhu}
\affiliation{National Astronomical Observatories, Chinese Academy of Sciences, Beijing 100012, China}

\author{Di Li}
\affiliation{National Astronomical Observatories, Chinese Academy of Sciences, Beijing 100012, China}
\affiliation{University of Chinese Academy of Sciences, Beijing 100049, China}

\begin{abstract}
The Search for Extraterrestrial Intelligence (SETI) attempts to address the possibility of the presence of technological civilizations beyond the Earth.
Benefiting from high sensitivity, large sky coverage, an innovative feed cabin for  China's Five-hundred-meter Aperture Spherical radio Telescope (FAST), we performed the SETI first observations with FAST's newly commisioned 19-beam receiver; we report preliminary results in this paper. Using the data stream produced by the SERENDIP VI realtime multibeam SETI spectrometer installed at FAST, as well as its off-line data processing pipelines, we identify and remove four kinds of radio frequency interference(RFI): zone, broadband, multi-beam, and drifting, utilizing the Nebula SETI software pipeline combined with machine learning algorithms.  After RFI mitigation, the Nebula pipeline identifies and ranks interesting narrow band candidate ET signals, scoring candidates by the number of times candidate signals have been seen at roughly the same sky position and same frequency, signal strength, proximity to a nearby star or object of interest, along with several other scoring criteria.  We show four example candidates groups that demonstrate these RFI mitigation and candidate selection. This preliminary testing on FAST data helps to validate our SETI instrumentation techniques as well as our data processing pipeline.

%To conduct a large SETI sky survey a large amount of %observation time is needed. It would be difficult to obtain %enough observing time at a large telescope such as China's %500-meter Aperture Spherical Radio Telescope (FAST) using the %normal time allocation mechanism. To get as much telescope %time as possible, we use a commensal observation program for %SETI. This means that we do SETI data analysis while the %telescope performs other observations. 

%In this paper, we first introduce our FAST SETI back end for %commensal observation and our pipeline for data processing. %We show the performance of our radio frequency interference %(RFI) removal and candidates selection program through %several hours of data from FAST. Candidates from the data %will not be discussed in details, because we just want to %verify the validity of our back end and pipeline in this %paper. We will collect more data during FAST's drifting scan %survey over the two to three years, and we hope to find %candidates with more evidence for ETI.  

\end{abstract}


\section{Introduction}\label{sec:intro}
The search for extraterrestrial intelligence (SETI), also known as the search for technosignatures \citep{tarter_2006,2018arXiv180906857W}, is a growing field in astronomy. This is partially due to the super-computer and big data revolution, machine learning technology, the privately-financed Breakthrough Listen Initiative, the thousands of recently discovered exoplanets, as well as the construction of new facilities, including FAST \citep{8331324}. 
\par
 “The probability of success is difficult to estimate, but if we never search the chance of success is zero” \citep{1959Natur.184..844C}.
 \cite{GrimaldiE9755} calculate the number of electromagnetic signals reaching Earth based on parameters from the Drake Equation.  \cite{2016JCAP...08..040L} calculate the relative formation probability per unit time of habitable Earth-like planets within a fixed comoving volume of the Universe. They found that life in the Universe is most likely to exist near $\sim$ 0.1 $M_{\odot }$ stars ten trillion years from now. \cite{2019MNRAS.485.5924L} study photosynthesis on habitable planets around low-mass stars to examine if this kind of planet can receive enough photons in the waveband of an active range of 400-750 nm to sustain Earth-like biospheres. 
 \par
 Radio SETI \citep{1959Natur.184..844C} is an important technique because Earth's atmosphere is relatively transparent at many radio wavelengths, and radio emissions have low extinction through the interstellar medium (ISM)\citep{2001ARA&A..39..511T,2013ApJ...767...94S}. Many radio SETI experiments have been conducted at Green Bank, Arecibo, Parkes, Meerkat, and several other single dish and array telescopes. Some recent examples are \citet{2013ApJ...767...94S}, \citet{2013ApJ...767...94S}, \citet{2018PASP..130d4502M}, \citet{2018PASA...35...41P}, \citet{Chennamangalam_2017}, and \citet{2017ApJ...849..104E}.   The first experiment for SETI with the Murchison Widefield Array (MWA), one of four Precursors for Square Kilometre Array (SKA) telescope, has an extremely large field of view \citep{2016ApJ...827L..22T}. The VLA SETI experiment by \cite{2017AJ....153..110G} implemented the search for artificial radio signals from nearby galaxies M31 (Andromeda) and M33 (Triangulum).  
\par
FAST, Earth's largest single-aperture telescope \citep{2000ASPC..213..523N,doi:10.1142/S0218271811019335}, has unique 
advantages for SETI observations.  FAST can observe declinations from $-14.3^{o}<\delta <+65.7^{o}$ (versus $+1.5^{o}<\delta < +38.5^{o}$ 
at Arecibo) due to its geographical location and active surface. FAST's sensitivity is $\sim 1800\ {\mathrm m^2/K}$ (Arecibo is 
about $1100$ $m^2/K$).  
 \par
This paper presents the first results of SETI at FAST.
The FAST SETI instrument was installed by the UC Berkeley SETI group in September 2018, and we have conducted preliminary commensal and targetted observations over the past year.  This paper is organized as follows: Section \ref{sec:2} describes the commensal observations at FAST and an overview of the SETI analysis pipeline. SERENDIP data acquisition, reduction, and analysis are described in Section \ref{sec:3}. Radio frequency interference (RFI) removal is discussed in Section \ref{sec:4}, and Machine Learning for radio frequency interference (RFI) removal and candidate selection are provided in Section \ref{sec:5}. Results is presented in Section \ref{sec:6}. Conclusion and Future Plans are in Section \ref{sec:7}. 

\section{Commensal observation for SETI at FAST}\label{sec:2}
Radio astronomy is a discipline that relies on observation. However, observing time on large telescopes is typically oversubscribed and often only one source can be observed at a time. To increase sky coverage, commensal observation \citep{BOWYER1983147} is being increasingly employed. During commensal observation, although the direction of the telescope is determined by the primary observer, secondary observers can receive a copy of the raw data in real time. This is the technique used for SETI at FAST.

%timely. Its basic working principle is: digital raw voltage data are stored in shared memory through a data acquisition system (the High Availability SHared Pipeline or HASHPIPE\footnote{\href{https://github.com/david-macmahon/hashpipe}, and then it is passed to a graphics processor (GPU) to form normalized spectra, finally record signals that exceed a power threshold of signal-to-noise ratio (SNR) (S/N = 30). These signals ("hits") are written to output files conforming to the FITS standard.} Figure \ref{fig:1} shows the architecture of FAST multi-beam digital back end.

The Chinese astronomical community has planned a drift-scan program covering the $57\%$ of the celestial sphere ($-14.3^\circ< \delta<+65.7^\circ$), called Commensal Radio Astronomy FAST Survey (CRAFTS)\citep{8331324}. CRAFTS plans to use more than 5,000 hours of telescope time. We plan to commensally analyze the sky survey data to find possible ETI candidate targets and to then do follow-up observations on these targets.

%The amount of raw data produced by these observations will be very large. Given a 19 beam receiver and a sampling rate of 1 billion samples per second (Gsps), the data produced per second can reach: 
%\[
%1 {\rm\ Gsps}\times 1 {\rm\ Byte/sample} \times 2 {\rm\ pol} \times 19 {\rm\ beams}= 38 {\rm\ GB/s}.
%\]
%Such volume of data is too much for available storage systems. An automated pipeline for data reduction, RFI removal, and candidate selection is very important. It is impossible to keep the raw data in storage systems for a long time and process them manually. So we use SERENDIP VI \citep{7436240,2000ASPC..213..485C}, a real-time data processing system, and Nebula \citep{korpela19b}, an off-line data analysis pipeline designed for use with SETI@home \citep{korpela19a} together with machine learning to reduce and analyze data. Figure \ref{fig:1} shows the overall data processing framework.

\begin{figure*}[tbp!]
\centering
\includegraphics[width = 1\linewidth]{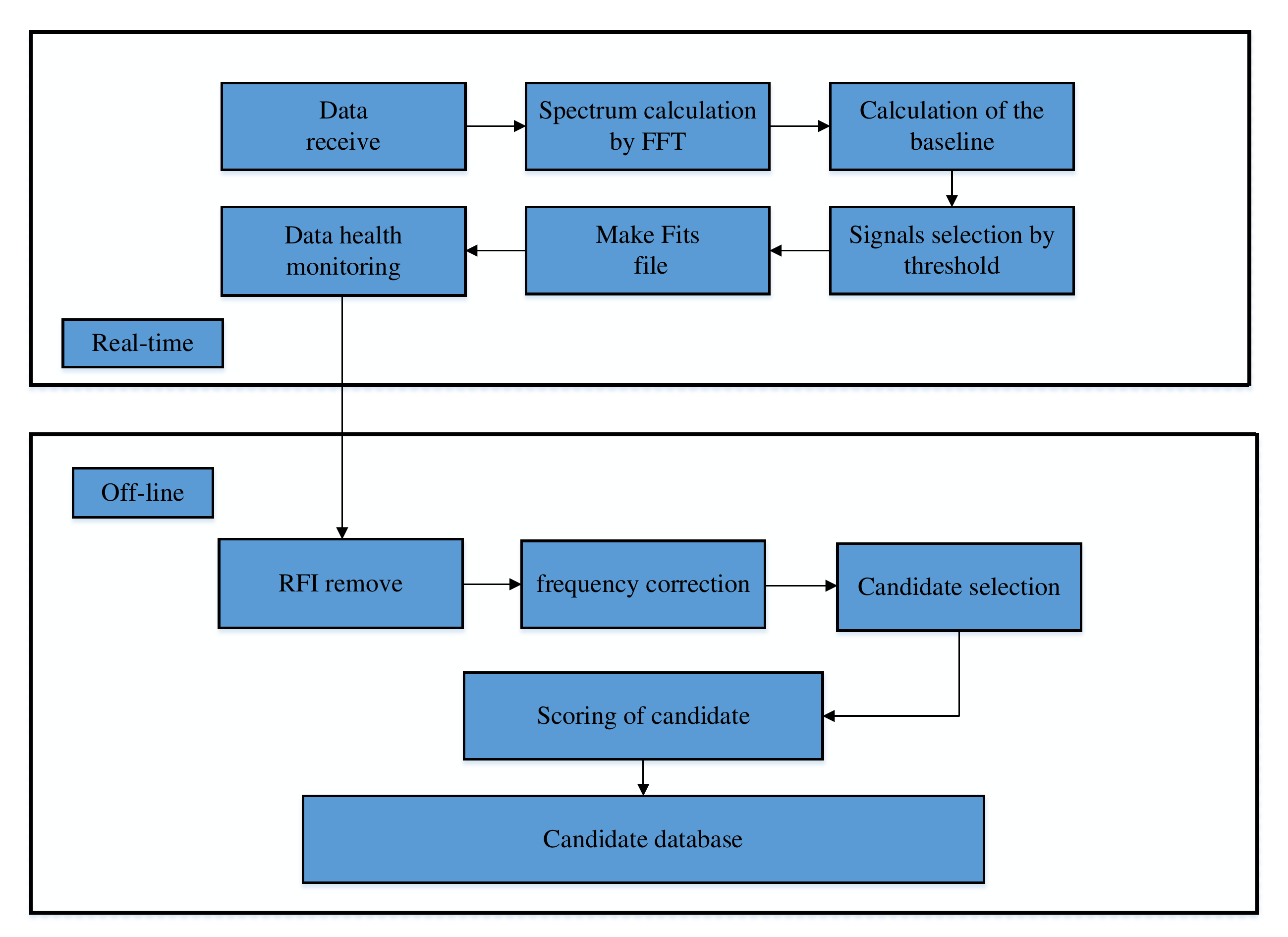}
\caption{Overall data processing framework. The whole pipeline consists of two main parts, real-time part and off-line part. In the Real-time part, we reduce the data according to the regulation we set and save into an ETFits file. In the Off-line part, we clean the data and select the candidates.}
\label{fig:1}
\end{figure*}

\section{SERENDIP-data acquisition, reduction and analysis}\label{sec:3}
%\subsection{SERENDIP VI}\label{sec:3.1}

The amount of raw data produced by these observations will be very large. Using the FAST 19 beam receiver and a sampling rate of 1 billion samples per second (Gsps), the data produced per second will be: 
\[
1 {\rm\ Gsps}\times 1 {\rm\ Byte/sample} \times 2 {\rm\ pol} \times 19 {\rm\ beams}= 38 {\rm\ GB/s}.
\]
Such a volume of data is too much for available storage systems. An automated pipeline for data reduction, RFI removal, and candidate selection is very important. We use SERENDIP VI \citep{7436240,2000ASPC..213..485C}, a real-time data processing system, and Nebula \citep{korpela19b}, an off-line data analysis pipeline originally designed for use with SETI@home \citep{korpela19a} together with machine learning to reduce and analyze data. SERENDIP VI is described in this section.  Figure \ref{fig:1} shows the overall data processing framework.

The SERENDIP VI system is a 128M channel spectrum analyzer, covering frequency bands from 1000 MHz to 1500 MHz with a frequency resolution of about 3.725 Hz. The system is composed of a front end, based on field programmable gate array (FPGA) systems and a back end based on GPUs, connected by a 10 Gbps Ethernet switch. 

\begin{figure*}[!htpb]
\centering
\includegraphics[width = 0.9\linewidth]{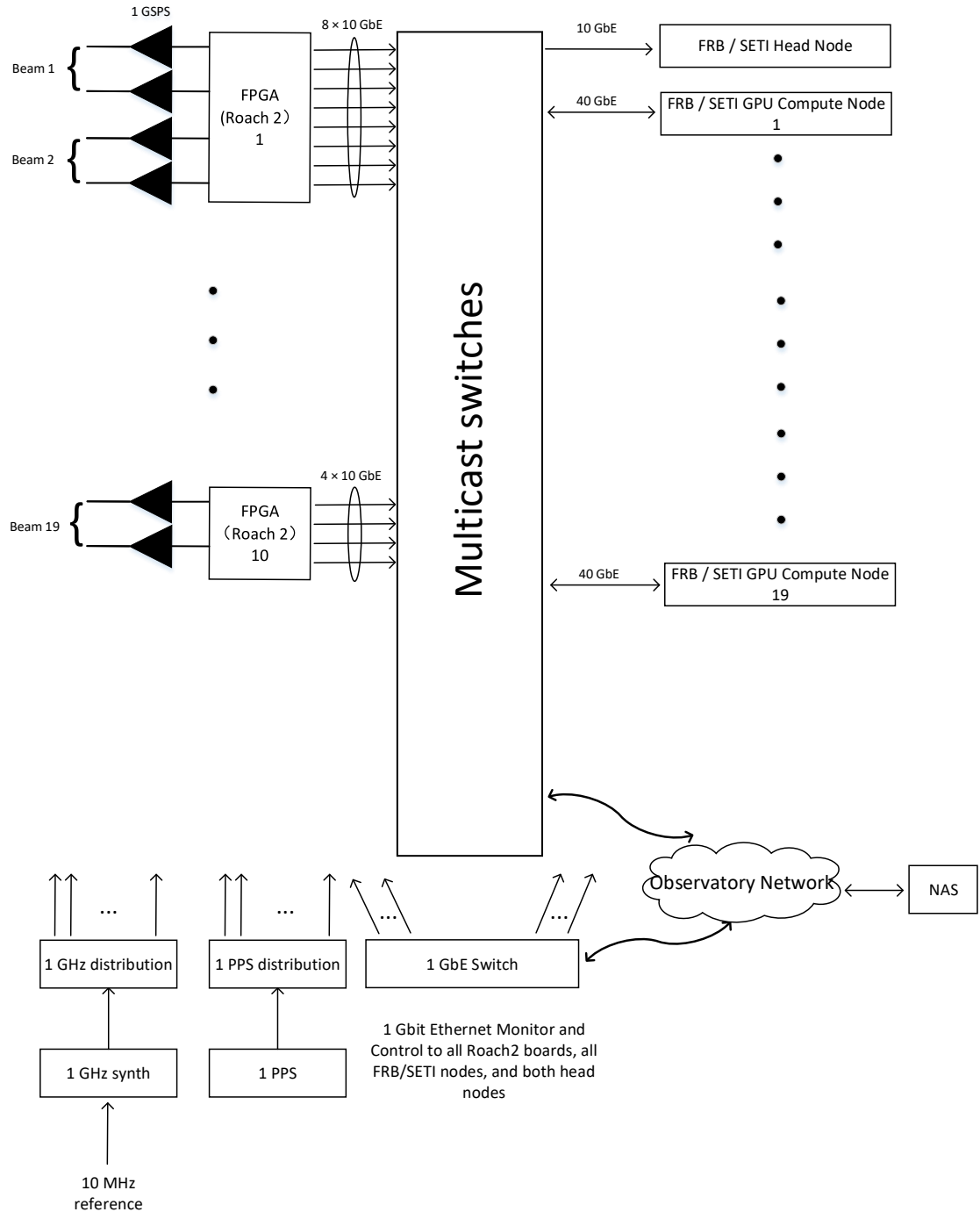}
\caption{Architecture of FAST multi-beam SETI instrument. 38 Analog signals (19 beams, each with two polarizations) are digitized and the digital samples are packetized by a bank of FPGAs (ten FPGAs are required for 19 beams at two beams per FPGA). Accumulated spectra for the FRB experiment are also formed and packetized at this stage.  Packets are then multicast through the 10GbE switch and received by GPU equipped compute nodes.  Each compute node handles one beam, computing high resolution spectra and identifying narrow band signals with powers significantly above the noise.  A head node provides control and monitoring.  Precise timing is achieved via a 10MHz reference and a 1 pulse per second signal. The entire instrument is also attached to the observatory network to allow remote development and operation.}
\label{fig:2}
\end{figure*}

The architecture of the FAST multi-beam SETI instrument is shown in Figure \ref{fig:2}. The front end performs analog to digital conversion resulting in 8 bit digital samples. These are packetized into 4 KB Ethernet packets and multicast across the network to the back end.   Each FPGA can process data from two beams, requiring 10 boards for the 19 beam feed array.  The FPGA system employed is CASPER ROACH2 (Hickish et al., 2016).   These are widely used and supported within the radio astronomy community.  In addition to sending raw samples to the SETI back end, the FPGAs form integrated power spectra for transmission to the Fast Radio Burst(FRB) back end. Multicast is employed so that identical data streams can be received and processed by multiple experiments. \par
At the back end, data reduction and analysis are performed on the GPU and the results, along with observatory meta-data (e.g. time, pointing, receiver status), are stored in files conforming to the FITS standard.  Each GPU based compute node handles one beam for both the SETI and FRB pipelines, requiring 19 compute nodes.  In addition to the compute nodes, there is a head node that handles monitor and control functions and hosts a Redis database used for cross-node coordination.

Precise timing is provided by the observatory's 10MHz reference and a one pulse-per-second (1 PPS) signal. Long term data storage and system backups are handled by the observatory's Network Attached Storage (NAS).
\par

The GPU processing pipeline consists of the following steps.

(1) The raw time-domain voltage data are copied to GPU memory and transformed into complex frequency domain data via cuFFT (a Fast Fourier Transform library provided by NVIDIA). 

(2) Based on the formula below:
\begin{equation}
{\mathrm Power =real^2+imag^2},
\end{equation}
we sum the real part and imaginary part of each channel in frequency domain to get the power spectrum.  This and subsequent steps are coded as calls to Thrust, NVIDIA's C++ template library.  

(3) The baseline of the power spectrum is calculated with respect to the local mean utilizing a sliding 8K spectral bin window.  The spectrum is then normalized with respect to this baseline.

(4) Finally, the normalized power of each channel is compared to a signal-to-noise (S/N) threshold value. Those channels exceeding the threshold (S/N $>$ 30) are recorded in the FITS file, including the signal time, frequency, detection power, mean power, telescope pointing, and other information. Each channel recorded is called a hit.

\section{Nebula for Radio frequency interference removal}\label{sec:4}
Radio frequency interference (RFI) removal has always been a crucial part of radio telescope data analysis. There has been much work done to date on the problem of RFI removal, including SumThreshold method \citep{2010MNRAS.405..155O}, Singular value decomposition (SVD), Surface fitting and smoothing  \citep{2007AN....328...68W} and Sky-Subtracted Incoherent Noise Spectra (SSINS) \citep{2019arXiv190601093W}.

Nebula is a complete off-time data analysis system, including data cleaning, RFI removal,  candidate selection, and scoring. Here we just give a brief introduction of RFI removal part with the example from FAST data. 

\subsection{Narrow-band RFI}\label{sec:4.1}
Narrow-band RFI is the most common kind of RFI coming from artificially engineered signals on the Earth, especially within the FAST electromagnetic environment. In Nebula, we separate the narrow-band RFI into two kinds. Narrow-band RFI that is stable in frequency (called zone RFI in Nebula) and Narrow-band RFI that drifts in frequency (called drifting RFI in Nebula). 

\subsubsection{Zone RFI}\label{sec:4.1.1}
\begin{figure*}[h!]
\centering
\includegraphics[width = 1\linewidth]{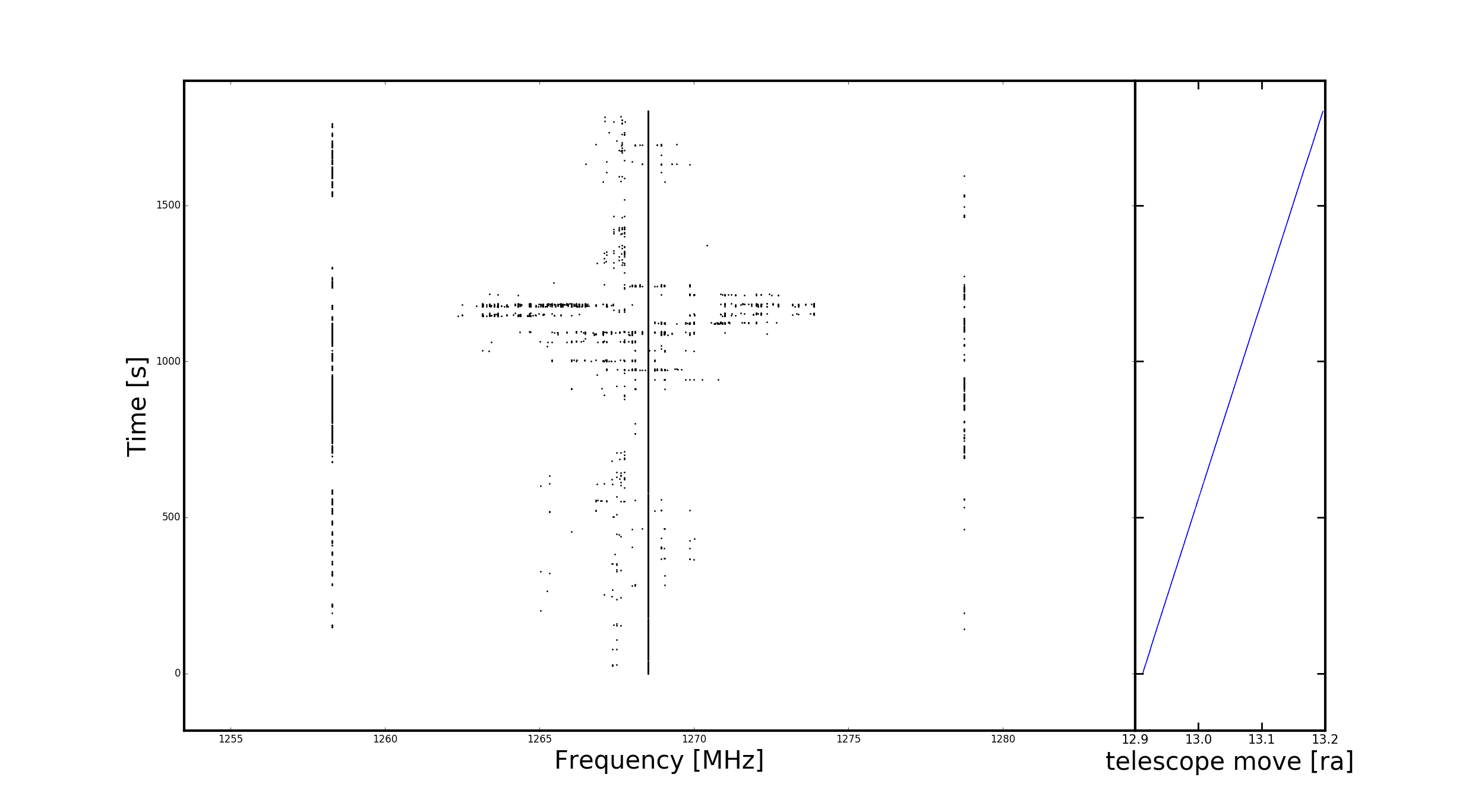}
\caption{Waterfall plot showing time versus frequency for a zone RFI and the right ascension of FAST in Equatorial coordinates. There is a typical zone RFI around 1268 MHz. Because they often contaminate the entire frequency channel, we call them zone RFI. Note, here and hereafter for all of the waterfall figures below, that the values of coordinates RA for telescope movement is a simulated information for a drifting observation. We give a explanation in Section \ref{sec:4.3}.}
\label{fig:3}
\end{figure*}

Zone RFI is narrow-band RFI exhibiting a stable frequency which persists throughout most or all of the total data set. These "zones" become global exclusion filters. There are many sources of such interference, including television, radio broadcasts, cell phone, and satellite signals. As shown in Figure \ref{fig:3}, compared with the distribution of the hits we expected, the zone RFI in the middle is more concentrated, forming a vertical narrow band. The existence of such RFI seriously affects the extraction of candidate targets. 

For zone RFI, we examine each frequency bin and calculate the number of hits in this frequency bin. If the number reaches above the threshold, all the hits in this bin can be marked as zone RFI and removed. The threshold is set by the Poisson cumulative probability function. The Poisson cumulative distribution is used to model the number of events occurring within a given time interval. Poisson cumulative distribution probability function can be expressed as:
\begin{equation}
F(x;\lambda) = \sum_{i=0}^{x}\frac{e^{-\lambda }\lambda ^{i}}{i!},
\end{equation}
where $\lambda$ is the mean number of hits in a bin, and x is the number of the hits. This function can give the probability of the number of hits in a bin. For the FAST data, we set the probability to be $1\times 10^{-7}$, so we can get a threshold value $x$ from the function.

\subsubsection{Drifting RFI}\label{sec:4.1.2}
\begin{figure*}[h!]
\centering
\includegraphics[width = 1\linewidth]{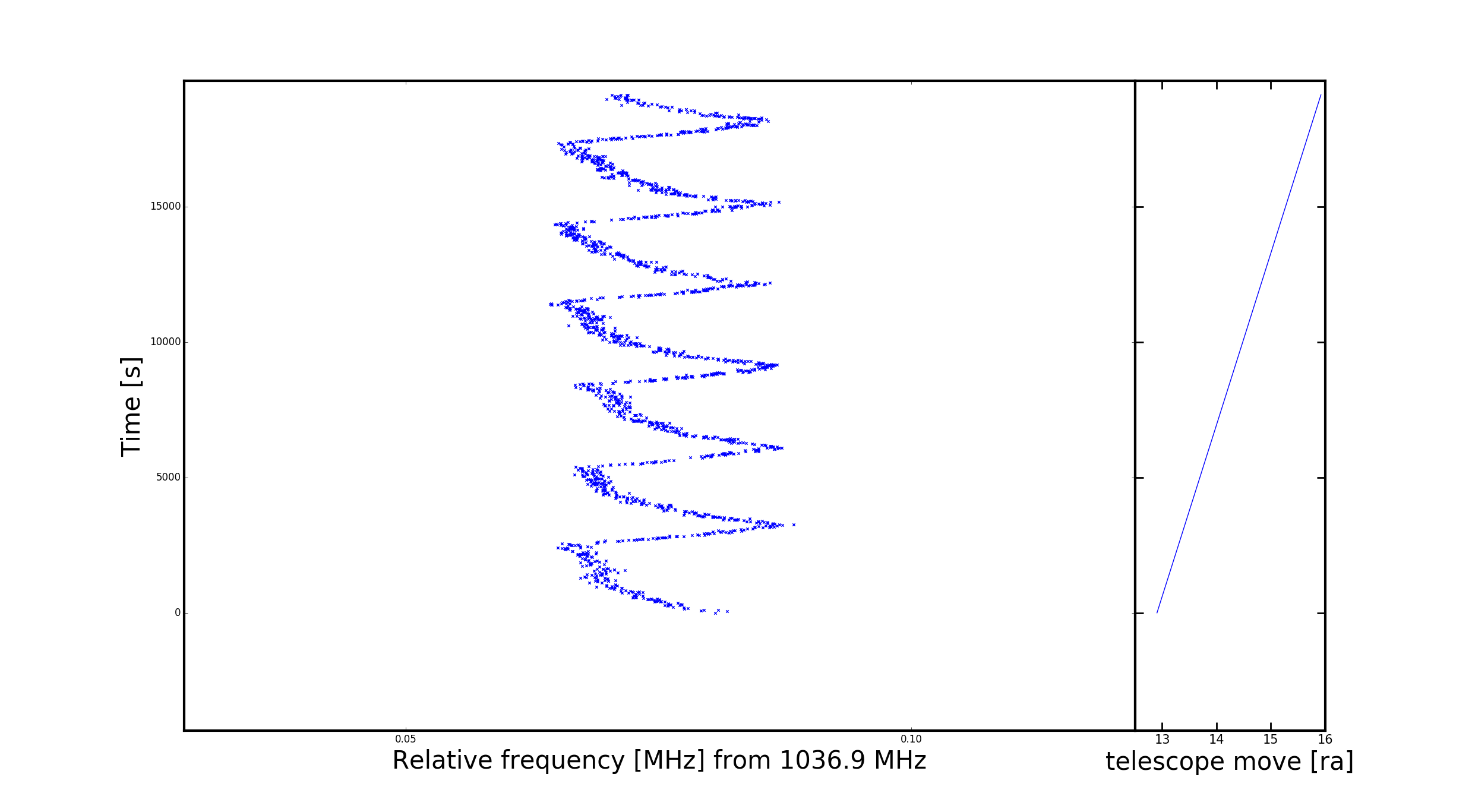}
\caption{Same as Figure \ref{fig:3} but for a drifting RFI.}
\label{fig:4}
\end{figure*}

Drifting RFI is narrow-band RFI that drifts in frequency, mainly coming from mobile devices. We can not find them with zone RFI algorithm, because the frequency of signals is changing over time. Figure \ref{fig:4} is a typical drift RFI from FAST data.

\begin{figure*}[h!]
\centering
\includegraphics[width = 1\linewidth]{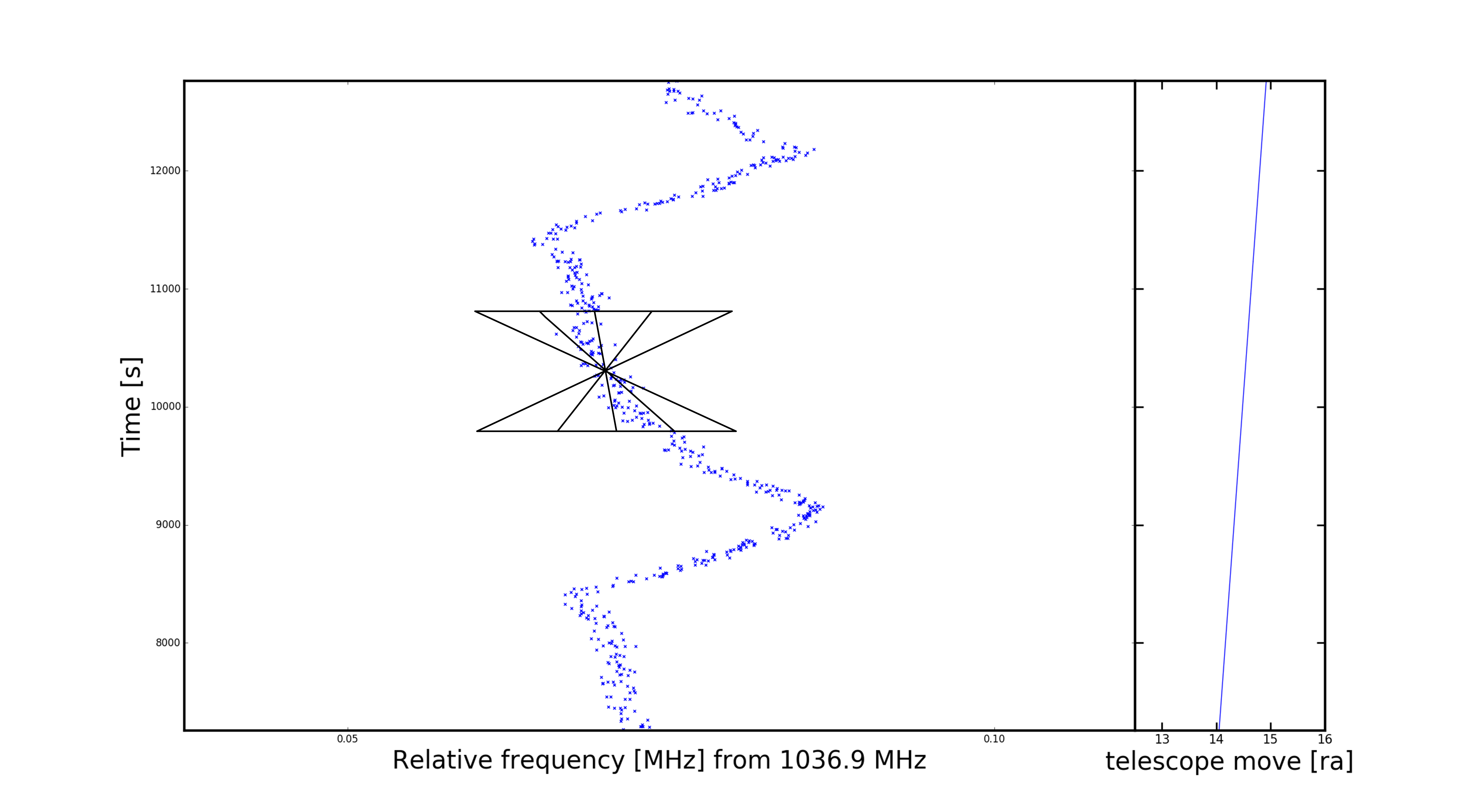}
\caption{Method of removing drifting RFI in waterfall plot of Figure \ref{fig:4}. For simplicity, we only show four bins of the whole 21 bins there.}
\label{fig:9}
\end{figure*}

Due to its changing frequency, we cannot simply remove it by frequency channel. In Nebula, we make two symmetrical triangles for each hit. As shown in Figure \ref{fig:9}, the shape of the triangle is determined by drift rate and time, which are set empirically. For FAST data, we set drifting rate to 20 Hz/s and $\Delta$t to 600 s. Then we can separate the triangle into 21 bins. If the number of signals in each bin and its opposite three bins is above the threshold, we mark all the signals in the bins as drifting RFI. The threshold is set in the same way as Section 4.1. 

\subsection{Multi-beam RFI}\label{sec:4.2}
When we use the multi-beams receiver such as the 19 Beams from FAST, we can identify signals that come from non-adjacent beams but with similar time and frequency. When a signal comes from a point in space, it can be received by one beam and maybe an adjacent beam. However, terrestrial RFI signals are often picked up in multiple beams simultaneously. In this algorithm, each hit has a time and frequency box. If the box of one hit is in another hit's box and hits are in non-adjacent beams, they are both marked as RFI.

\subsection{Pipeline of RFI mitigation}\label{sec:4.3}
\begin{figure}[tbp]
\centering
\includegraphics[width = 1\linewidth]{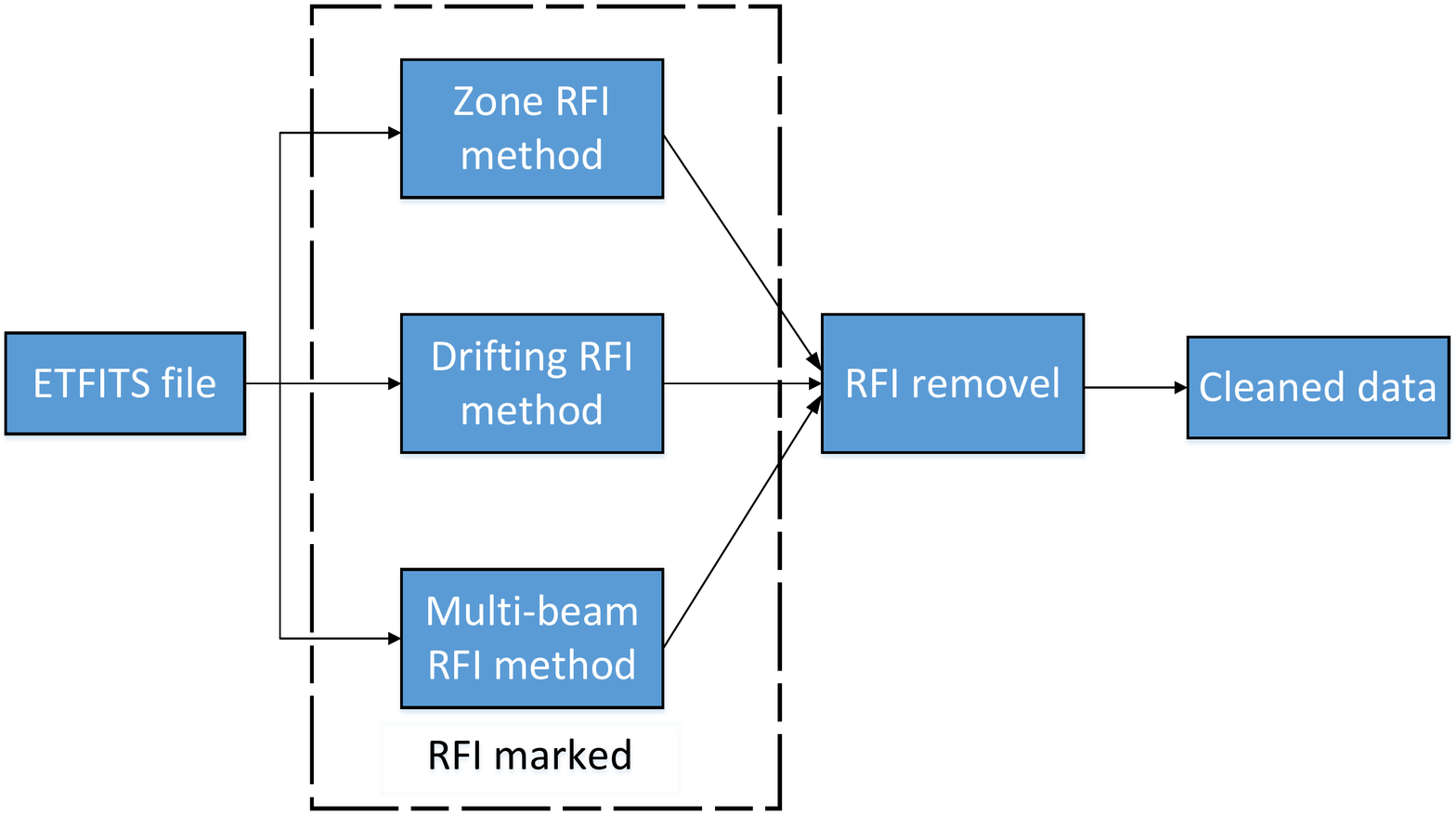}
\caption{Block diagram of Nebula RFI removal. The full set of raw data go through each of three RFI identification methods. Any hit that is marked by a least one of these methods is removed. All remaining hits constitute the set of clean data that is passed on to candidate selection.}%Block diagram of Nebula RFI removal. The full set of raw data go through each of three RFI identification methods. Any hit that is marked by a least one of these methods is removed. All remaining hits constitute the set of clean data that is passed on to candidate selection.}
\label{fig:20}
\end{figure}

Section \ref{sec:4.1} and \ref{sec:4.2} demonstrate three kinds of RFI and their remove methods. In Nebula, we don't remove these three kinds of RFI separately, because narrow-band RFI may probably be multi-beam RFI. We just mark the hit instead of removing it when the hit is selected by any one of three methods in Nebula. Finally, the hit with one or more RFI labels will be removed after going through all three kinds of methods. Figure \ref{fig:20} shows the processing pipeline of Nebula. 
{\bfseries   When the data was taken for this paper, real time pointing information was not available from the FAST telescope. 
For this early test data, telescope position information was calculated and merged with the SETI data in post processing. 
We cross checked our approximate calculated positions with the more accurate positions provided to us later by the FAST telescope pointing system -- our position errors are under 0.014 degree, which is adequate for testing the Nebula data analysis pipeline, although not ideal for the upcoming sky survey.   
We expect the pointing errors in the upcoming sky survey will be significantly lower, as our SETI spectrometer will have access to accurate real time telescope pointing data, and the spectrometer will merge this pointing data with the science data before it's recorded to disk.}

%{\bfseries Note that all the RA and DEC appeared in the paper is not provided by FAST. FAST can not provide real-time telescope pointing information when we took data. So we add the calculated RA and Dec and merged it data in real-time in order pipeline to find RFI and candidates of ETI signals. During a drifting scan, telescope’s movement is very simple, fixed Dec and RA synchronized with the rotation of the earth. So in all of the waterfall plots of this paper, we only show the calculated RA which allows the pipeline to identify RFI and candidates. We cross-check the Ra and Dec calculated method with the pointing information provided by FAST according to the feed cabin position file and coordinate transformation. The result is shown in Figure \ref{fig:Ra_residual}. During the whole 5 hours observation, the deviation is about 50 arc seconds, with an average deviation of 0.0028 arc seconds per second. The reason we did this is because the telescope pointing information calculated by ourselves can let us know the sky angle that the telescope moved for our RFI removal and target selection pipeline. For example, if the signal keeps appear and the telescope is moving, we will mark it as RFI in the RFI removal part. In the pipeline, the sky region that the telescope actually points to is not regarded. When our pipeline gets the candidate target, we will manually verify the exact location of the candidate target for scientific analysis. }

\begin{figure}[tbp]
\centering
\includegraphics[width = 1\linewidth]{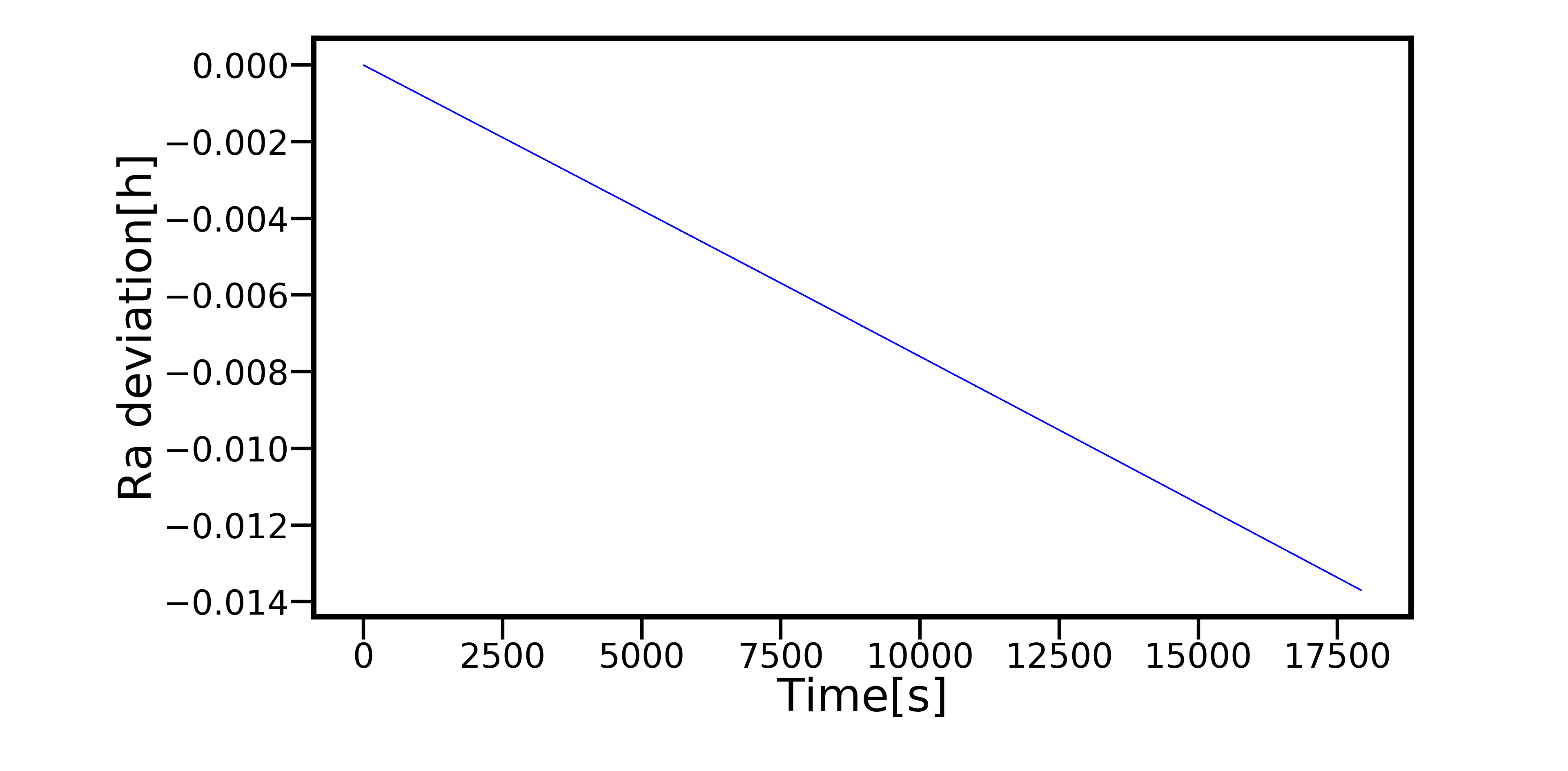}
\caption{Deviation between our calculated right ascension and the coordinates provided by FAST. The total deviation over five hours is about 50 arc seconds, with an average deviation of 0.0028 arc seconds per second.}%Ra deviation between our calculated Ra and Ra provided by FAST. The total deviation of five hour is about 50 arc seconds with an average deviation of 0.0028 arc seconds per second.}
\label{fig:Ra_residual}
\end{figure}

%\subsection{Broad-band RFI}
%Broad-band RFI do not often appear, but once present, it can contaminates many frequency bin and signals. It is mainly from natural lightning, sparks, or power transmission cables on the earth's surface. Here we do not remove this kind of RFI in nebula for FAST data but do it using Machine learning in Section \ref{sec:5}.

\section{Machine learning for RFI removal and candidate selection}\label{sec:5}
It should be noted that we employ the traditional assumption that advanced life wishing to be detected at interstellar distances will use narrow-band microwave emissions as narrow-band signals are most easily distinguished from emissions produced by natural astrophysical sources of microwave emission.  The narrowest known natural sources, astrophysical masers, have minimum frequency widths of about 500 Hz \citep{1987MNRAS.225..491C}. Thus we primarily focus on searches for narrow-band signals from ETI.

\subsection{Machine learning for RFI removal}
Our Nebula pipeline can remove most of the RFI. Normally, $>99\%$ of RFI can be removed, but there are still some atypical RFI left. Two examples are narrow-band RFI and broad-band RFI. We are unable to detect all narrow-band RFI, because at times the power is below our threshold. It is easy to find broad-band RFI if the bandwidth is very large.  If the bandwidth is less than several MHz, it is much more difficult and our traditional methods do not detect them.

\begin{figure}[tbp]
\includegraphics[width = 1\linewidth]{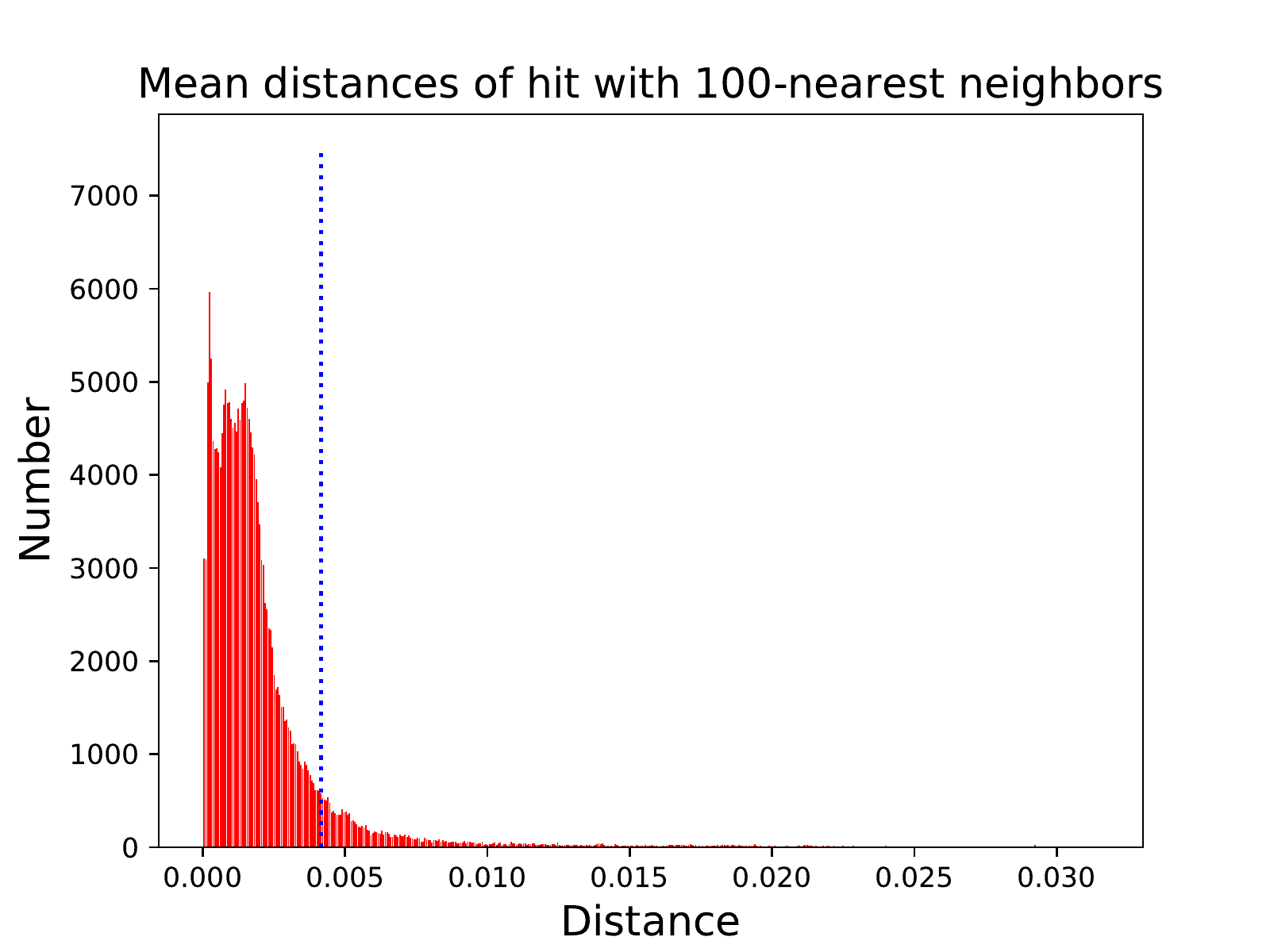}
\caption{Histogram of the number of hits versus mean distance. The vertical line is our chosen RFI removal threshold. Most hits are on the left side of the line with low mean distances, which means they come from big clusters, and can be removed. Distance shown in the graph is the mean distance between each hit and its nearest 100 hits in the waterfall plot showed before, expressed by time interval and frequency interval, i.e., $\sqrt{\Delta t^{2}+\Delta f^{2}}$.}
\label{fig:5}
\end{figure}

\begin{figure}[tbp]
\centering
\includegraphics[width = 1\linewidth]{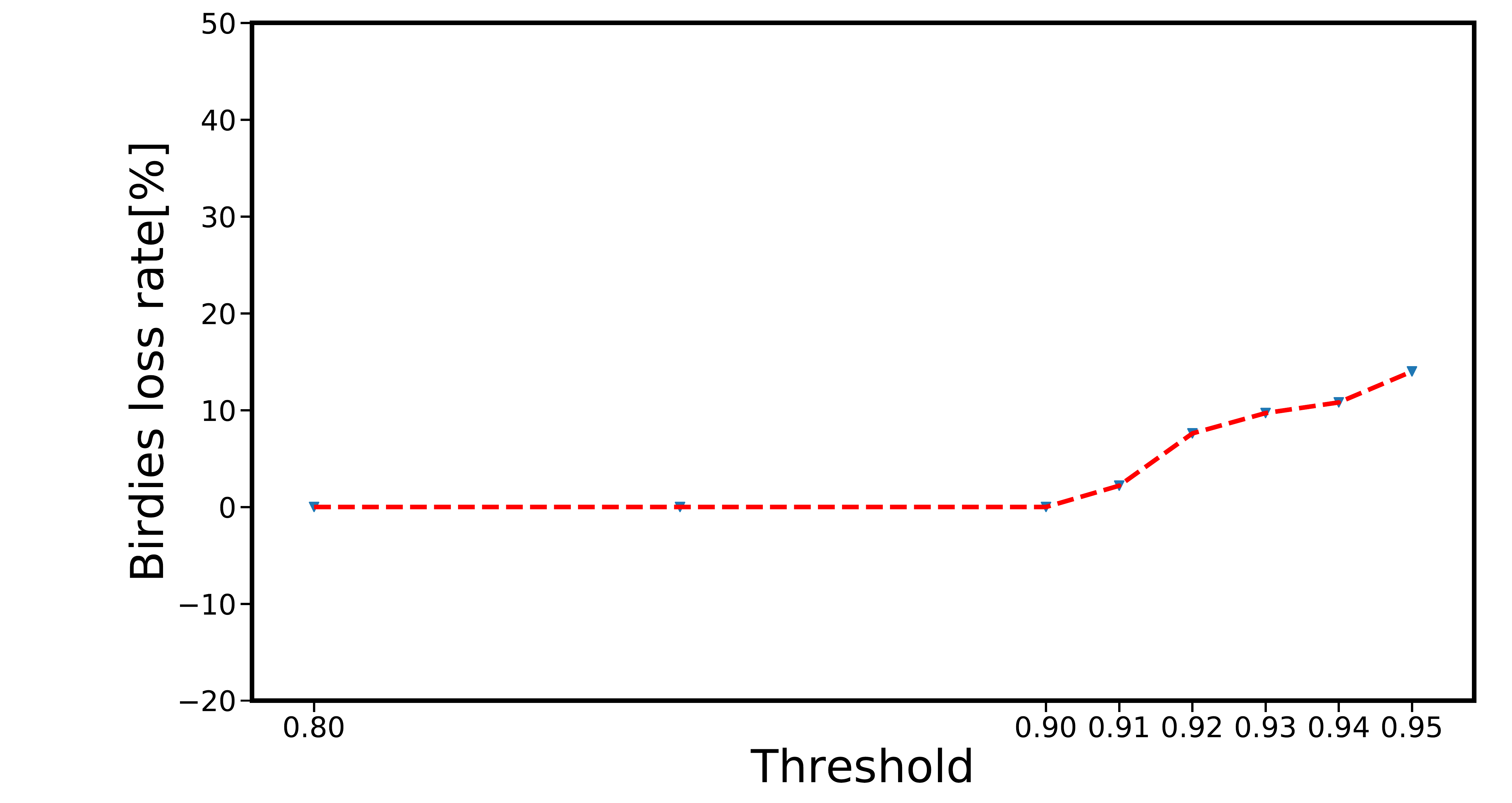}
\caption{Birdies loss rate curve. The experiment is based on the FAST data with birdies injected. As shown in the figure, birdies loss begins at a threshold of 91$\%$, so in order to reduce the possibility of potential signal loss, we choose the threshold of 90$\%$.}
\label{fig:21}
\end{figure}

Of course, these RFI signals still have characteristics in common: such emissions cluster in time and frequency on specific scales. The ETI signals typically would not form a large cluster on time and frequency scales, being narrow in frequency and of no longer duration than the telescope observed a point in the sky in the drift scan.  We use the k-Nearest-Neighbor(KNN) algorithm to find the nearest 100 hits for each hit and calculate the mean distance, as was first applied by \citet{vishala}. Figure \ref{fig:5} plots the histogram of hits' mean distances for the 100 nearest hits. The blue line is the upper RFI threshold, which is based on removing a specified percentage (90$\%$) of the total hits. Hits below this threshold are presumed to be RFI events.  We tested our threshold choice using experiments conducted to test the percentage of simulated "birdie" signals lost due to RFI removal at different threshold values. This can be seen in Figure \ref{fig:21}. The threshold value of 90\% conserved the most birdies, which maximizes the probability of conserving true ETI signals while minimizing the unremoved RFI. See Section \ref{sec:birdies} for more discussion of birdie generation.

\begin{figure*}[tpb]
\includegraphics[width = 1\linewidth]{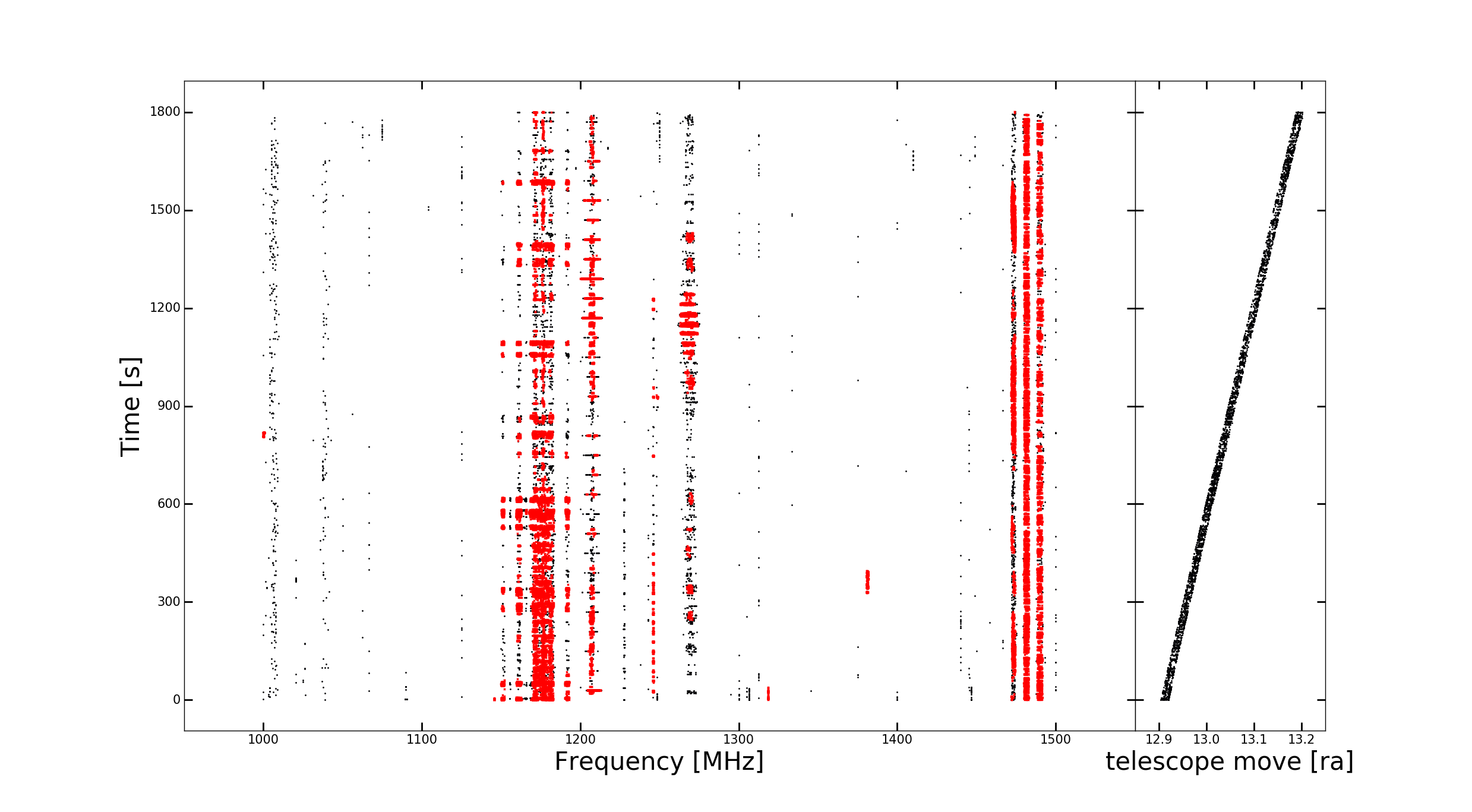}\\
\includegraphics[width = 1\linewidth]{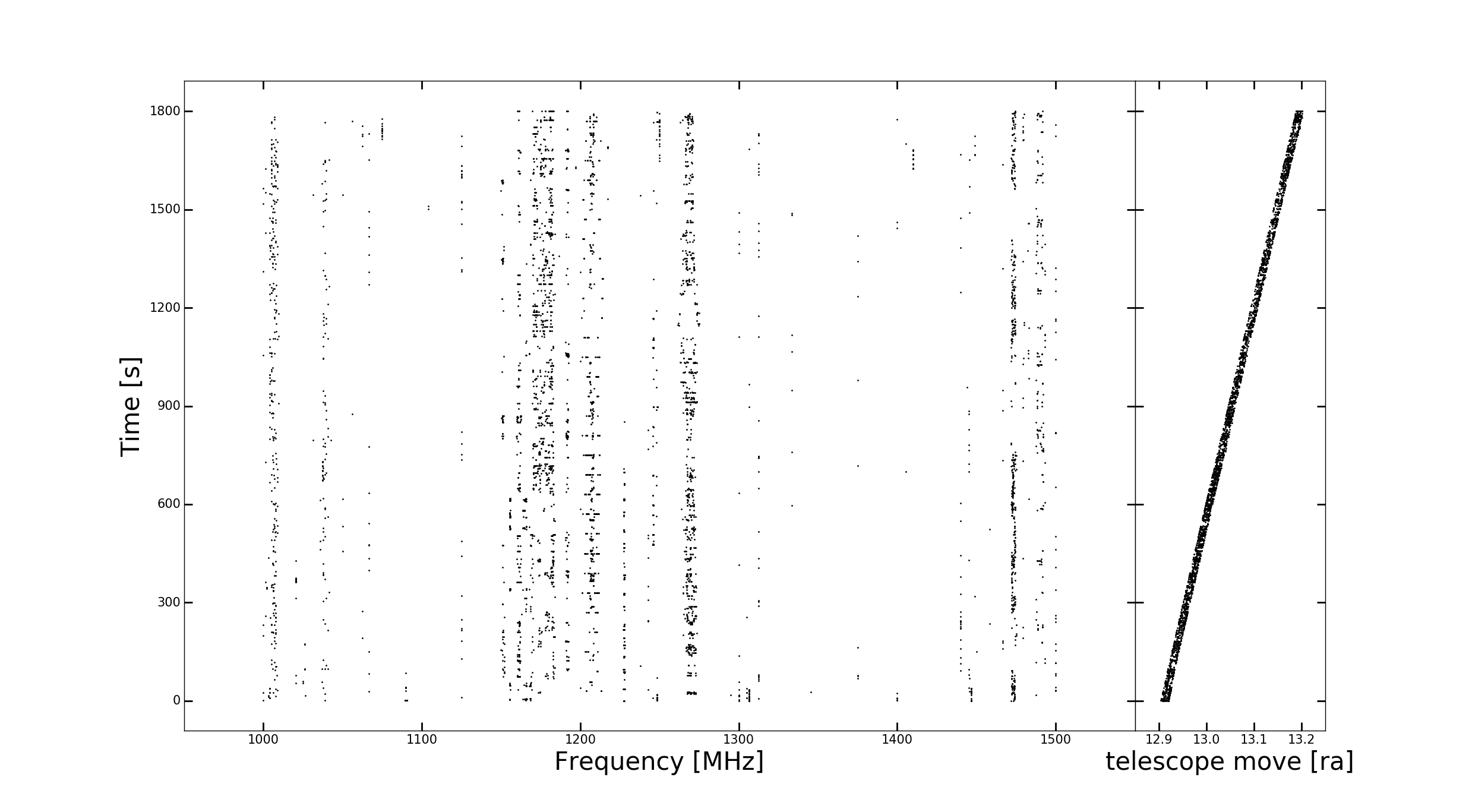}
\caption{Waterfall plot before (top panel) and after (bottom panel) RFI removal through machine learning. Top panel shows data including RFI, with the RFI marked by KNN marked in red. The bottom panel shows hits after KNN RFI removal. Although there is still part of RFI left in the bottom panel, KNN has removed more than 90$\%$ of RFI left by Nebula. The remaining hit are spare enough not to confuse for our candidate selection pipeline.}
\label{fig:6}
\end{figure*}

\begin{figure*}[tbp]

\includegraphics[width = 1\linewidth]{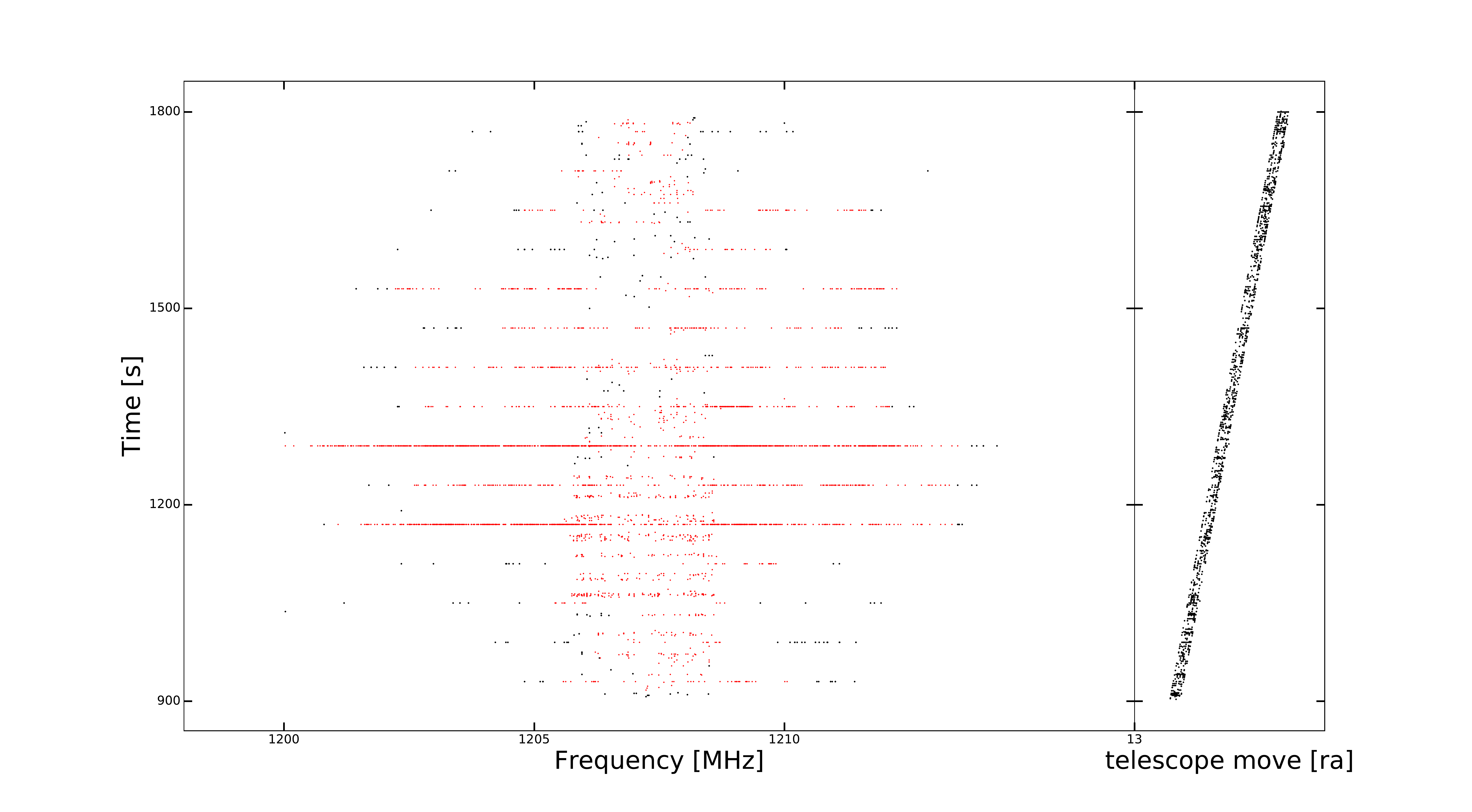}\\
\includegraphics[width = 1\linewidth]{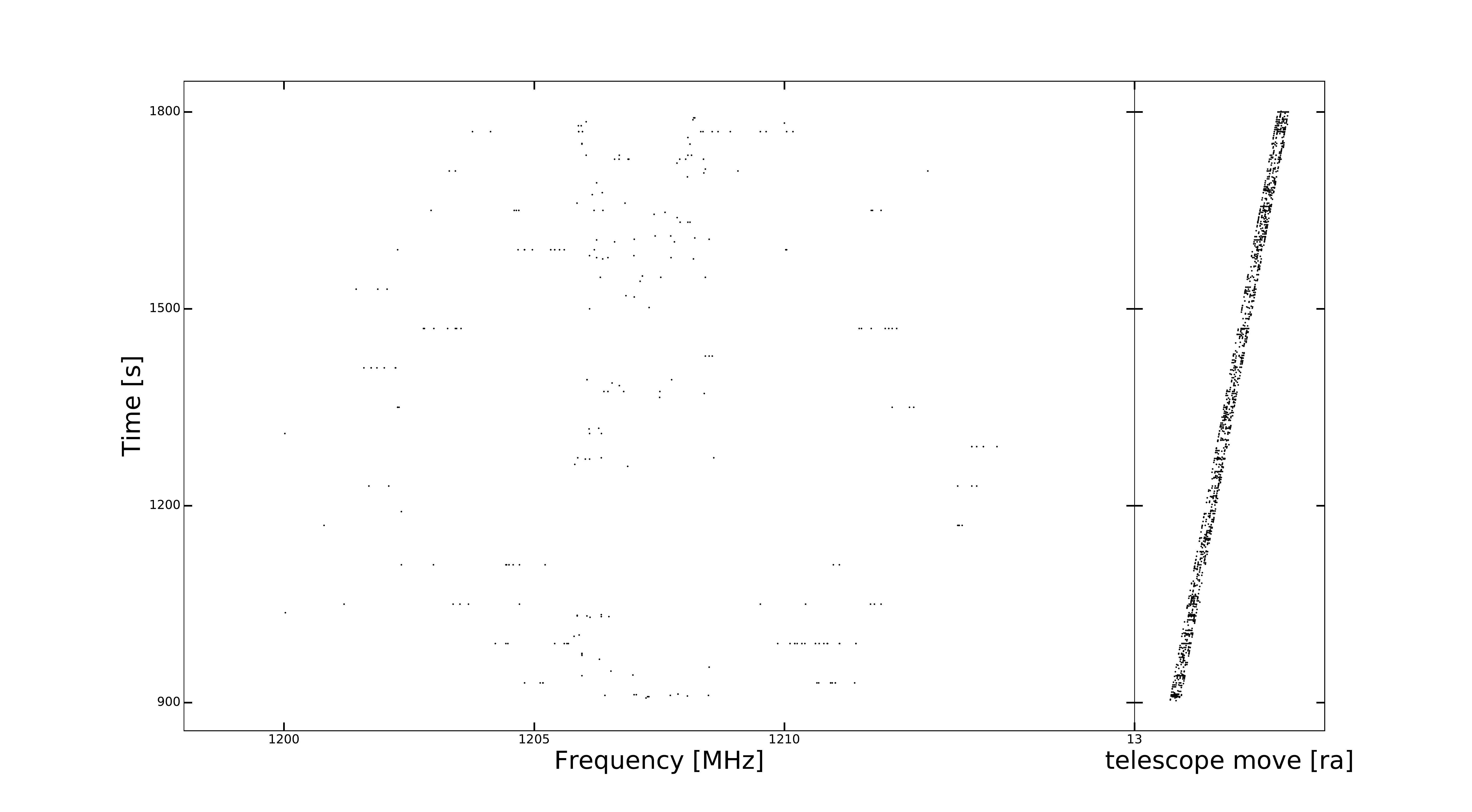}
\caption{Example of removing broadband RFI using machine learning. Broad-band RFI does not often appear, but once present, it can contaminate many frequency bins and signals. It is primarily due to lightning, sparks, or power transmission cables near the earth's surface. Top panel shows data including broad-band RFI, which are in red, while the bottom panel shows data after KNN RFI removal.}
\label{fig:8}
\end{figure*}

Figure \ref{fig:6} is a waterfall plot before and after using machine learning to remove RFI. Figure \ref{fig:8} is an example of removing broad-band RFI. The KNN algorithm can effectively remove much of the broad-band and narrow-band RFI left by Nebula. Following this removal the data are clean enough for selecting candidates.

\subsection{Machine learning for candidate selection}
After RFI removal, we apply a Density-Based Spatial Clustering of Applications with Noise (DBSCAN) \citep{Ester:1996:DAD:3001460.3001507}, a density clustering algorithm, to find candidate clusters. 

\begin{figure}[tbp]
\centering
\includegraphics[width = 1\linewidth]{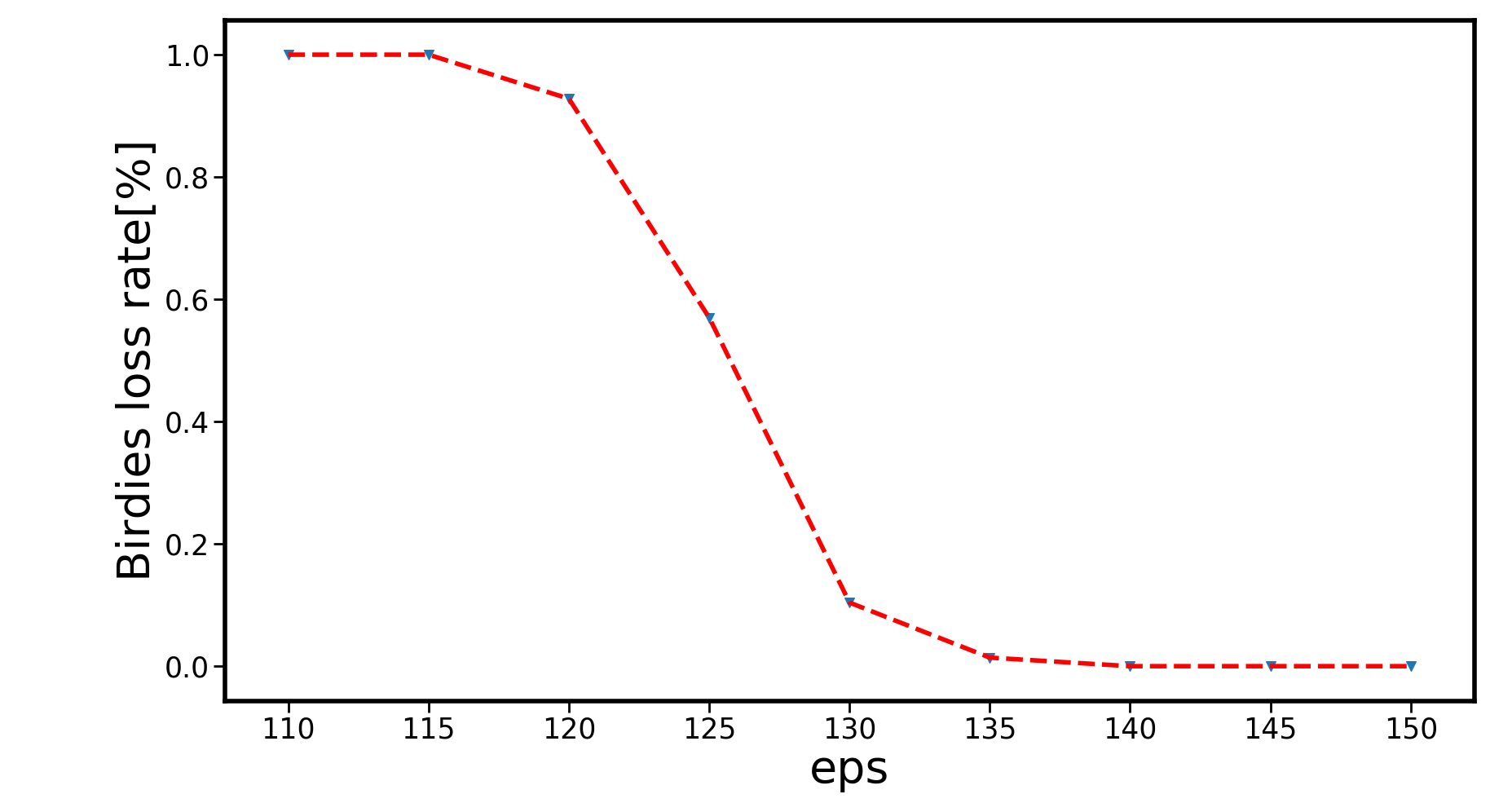}
\caption{Eps test curve. The experiment is based on the FAST data with birdies injected. As shown in the figure, birdies loss rate decreases from EPS equal to 115 and drops to 0 when EPS equal to 140.}
\label{fig:22}
\end{figure}

DBSCAN is a clustering algorithm based upon data density. It separates the data into three kinds of points: ``core points," ``adjacent points," and ``noisy points." ``Core points" are those with more than a predefined number of points within a specified radius, while ``adjacent points'' have fewer points nearby but belong to a core point. ``Noisy points'' neither have enough points around nor belong to a core point. DBSCAN does not divide data into parts but identifies tight clusters in any shape against the background. All the noisy points form a background group which then can be discarded. There are two main parameter:eps and Nmin. Nmin stands for minimum cluster population. We set this paramter to 5, because we don't want to miss candidate groups with only a few points. For eps which stand for maximum distance from the core point, we used a set of experiments to choose the most suitable value. The experiments result are shown in Figure \ref{fig:22}. When eps exceeds 140, the birdies loss rate drops to 0. We set eps to 145, because ETI signals will probably be a smaller cluster than our birdies during a drifting observation.
 
For each cluster, we apply two rules to distinguish RFI from non-RFI clusters.

\begin{enumerate}
\item We calculate the sky angle between hits in a cluster, and determine whether the cluster subtends less than 1.5 times the receiver beam's width. We use a larger than unity width because some  extraterrestrial signals could be received by a beam and an adjacent beam simultaneously.

\item We calculate the duration and bandwidth of the cluster. We select clusters with narrow bandwidth and a duration less than few tens of seconds.
\end{enumerate}

These two rules define the expected characteristics of extraterrestrial signals as they would appear during drift-scan sky survey observations. The final step is that we save all the candidates by location pixel. Pixels are defined to be small rectangles that divide the celestial sphere. All the candidates selected by the pipeline will be be saved by pixels in order to further analyze the candidate targets of the same sky position.

\section{Analysis and results for FAST data}\label{sec:6}

\subsection{FAST data}
\begin{figure}[tbp]
\centering
\vspace{-0.1cm}
\includegraphics[width = 1\linewidth]{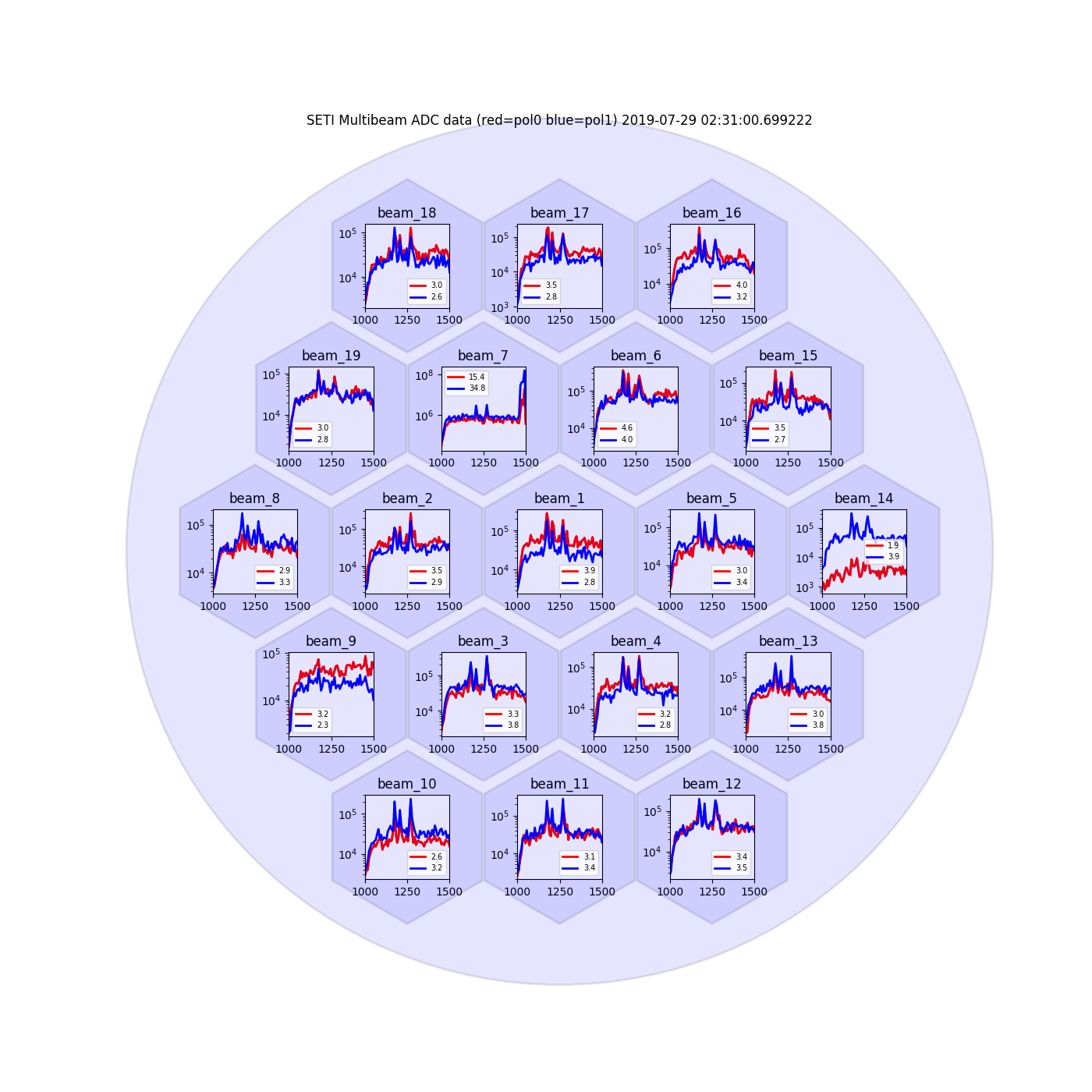}
\caption{SETI multibeam monitor. There are 19 plots in the figure stand for 19 beams.  Each plot has two lines. The line in red represents the power spectrum of polarization 0 while the line in blue represents the power spectrum of polarization 1.}
\label{fig:10}
\end{figure}

Our data were collected during a drift-scan survey performed by FAST during commissioning in July, 2019. To ensure beam health and data integrity, we employ a system health monitor. The monitor (Figure \ref{fig:10}) shows a coarse (2048 point) power spectrum (with RMS deviation) for each of 19 beams in both polarization at each beam's location in the focal array.   We update the power spectrum and the RMS in the plot windows every five minutes to monitor beam health. 

The data from each beam are saved into an FITS file separately. The size of each file is roughly one gigabyte and records a table of data for each hit received. Each hit contains 14 values, including time, position, beam number, power, SNR, channel number, and frequency.  Telescope information can be stored in header fields. This format is known as ETFITS and has been utilized by recent SERENDIP projects.

\subsection{Data processing\label{sec:birdies}}
\begin{figure*}[tbp]
\centering
\includegraphics[width = 1\linewidth]{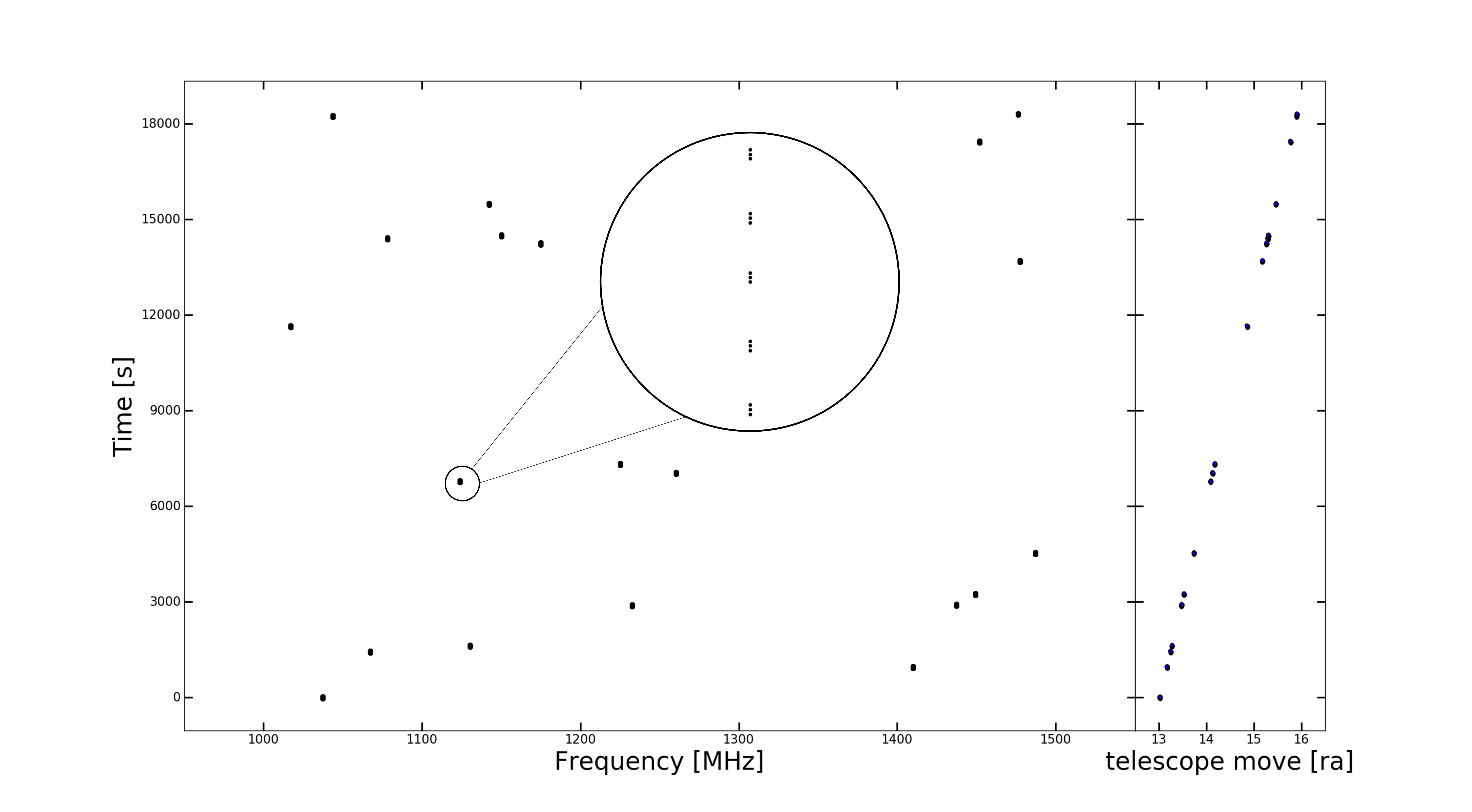}
\caption{Birdies generated for FAST data. Every black spot is a group of birdies. We zoom in one of them where details are shown in the larger circle. Each group contains several small groups that depend on how many beams go through the birdies.}
\label{fig:12}
\end{figure*}

In order to verify the validity of our pipeline, some artificial candidate targets, called ``birdies", are added to our data. We randomly generate some signals along the moving trajectory of beam one, and if other beams go through the same position, we add more signals with the same frequency into that beam. Birdies generated are shown in Figure \ref{fig:12}, which contains 20 groups and 294 signals.

\begin{figure*}[tbp]
\includegraphics[width = 1\linewidth]{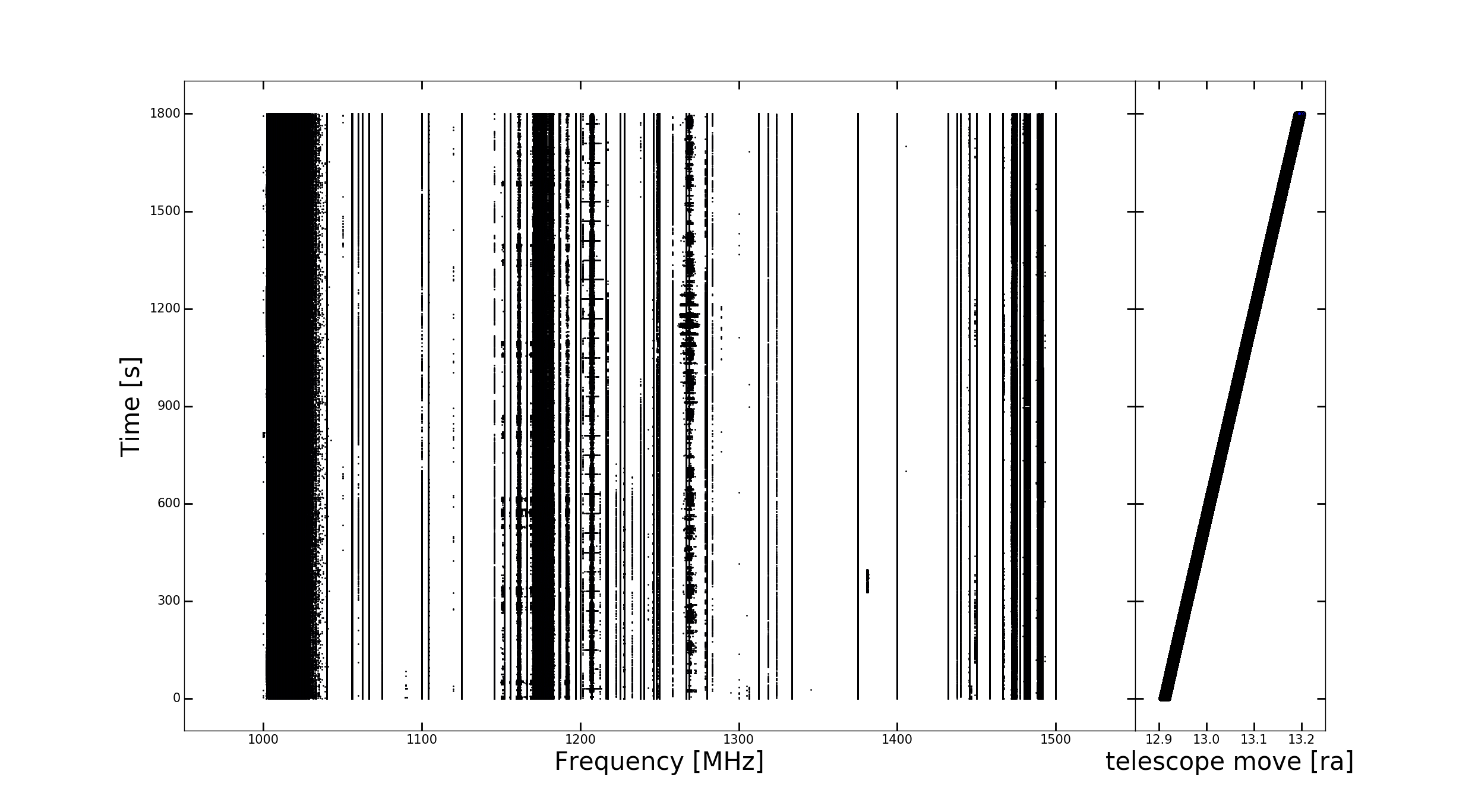}
\\
\includegraphics[width = 1\linewidth]{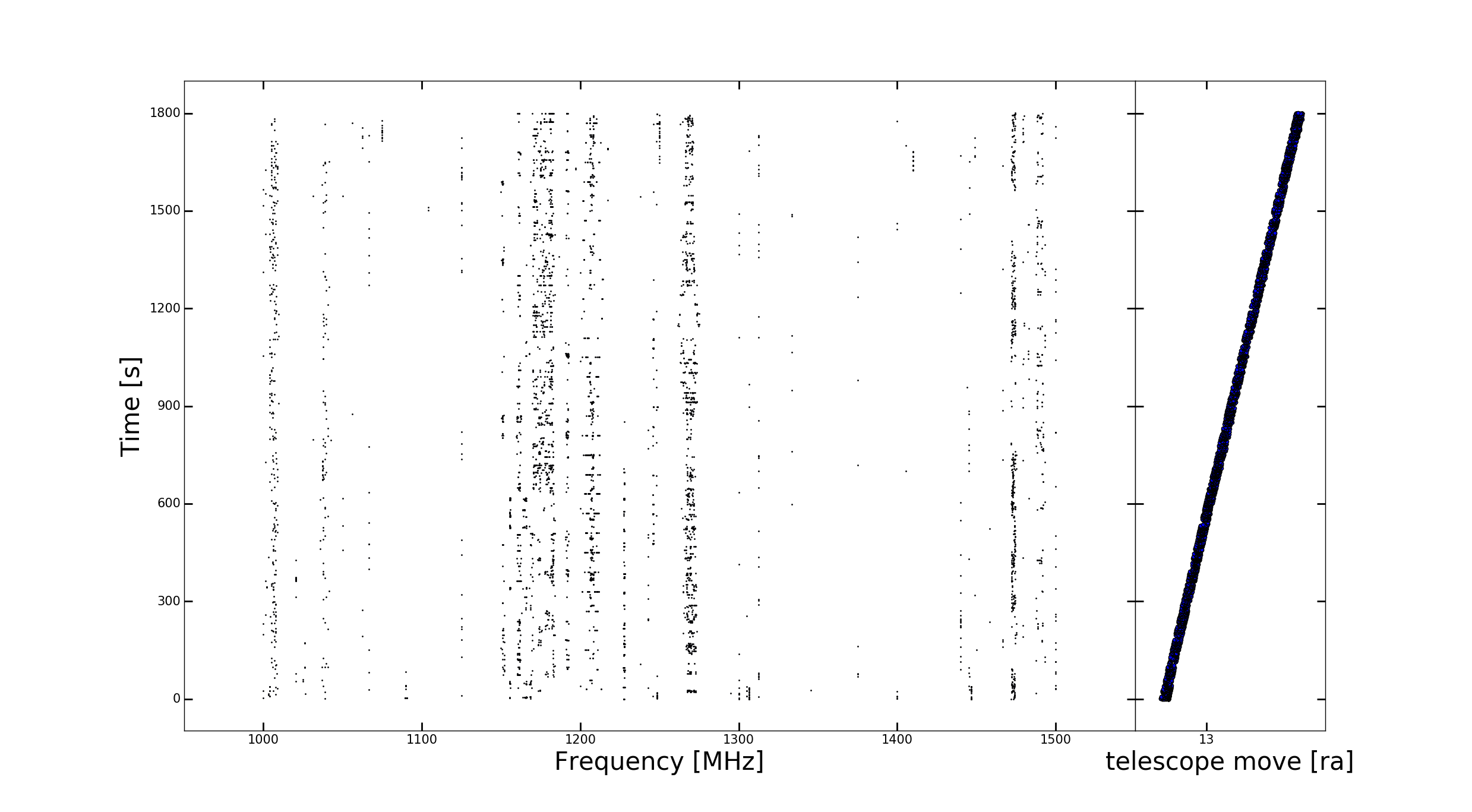}
\caption{RFI excision of FAST data. The top panel shows unprocessed data while the bottom one shows processed data. We can see directly that most of the RFI are removed. 98.1976$\%$ are zone RFI, denoted by the vertical bars in the top panel. Here note that we only show data spanning 1800 seconds in this figure.}
\label{fig:13}
\end{figure*}

After adding birdies, we use Nebula and the KNN pipeline to remove the RFI. 99.9063$\%$ of hits are removed by our pipeline while only 5.1020$\%$ of birdies were removed. Note that we make a temporary change to Nebula. To get an ideal velocity, we first remove zone RFI and then remove other kinds of RFI simultaneously. Details are shown in Table \ref{tab:5}. Most of the RFI in FAST are removed by the zone RFI algorithm. In the future, we will try to find the source of zone RFI and eliminate them earlier in the data acquisition process. This would make our data set smaller and cleaner, which could increase the probability of finding an ETI signal. We track the birdies removed by Nebula and find that they come from the same DBSCAN group and are heavily polluted by ambient RFI. Figure \ref{fig:13} shows the data before and after our pipeline.  We are convinced our pipeline effectively removes most of the RFI and protects most of the birdies and potential candidates.

\begin{center}
\begin{table}[!htpb]
    \setlength{\leftskip}{-20pt}
		\resizebox{0.5\textwidth}{!}{
			\begin{tabular}{lcccc}
				\hline
                \hline
				Type of RFI & Zone RFI & Drifting RFI & Multi-beam RFI & Total  \\
				\hline
				Number of bytes(MB) & 82524.9 & 735.2 & 700.9 & 83961 \\
				Percentage($\%$) & 98.1976 & 0.8748 & 0.8340 & 99.9063 \\
				\hline
			\end{tabular}
		}
		\\
		\caption{ Ratio of each kind of RFI}
		\label{tab:5}
	\end{table}
\end{center}

\subsection{Candidate selection}
\begin{figure*}[tbp]
\centering
\includegraphics[width = 1\linewidth]{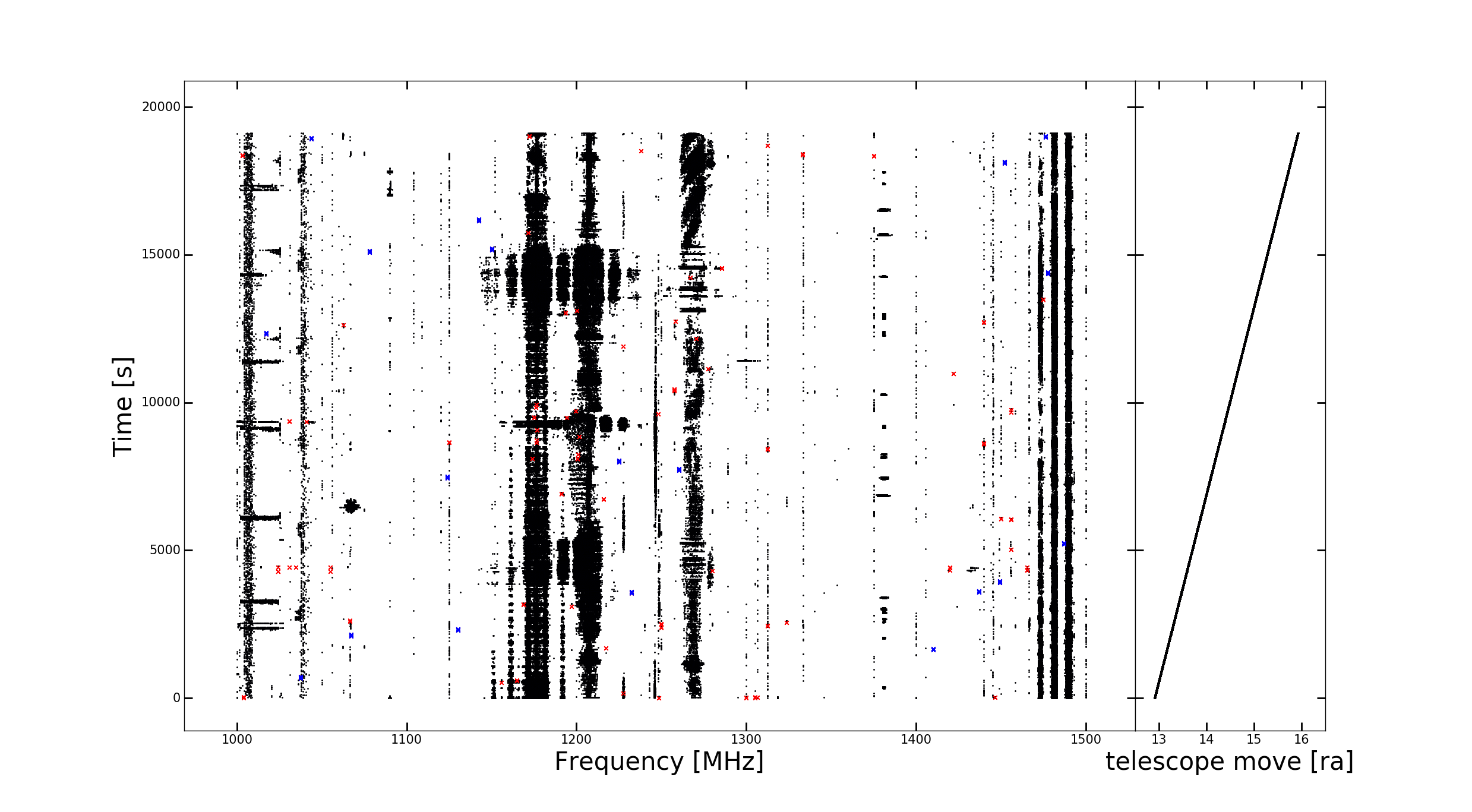}
\caption{Candidates found in FAST data. In the figure, Blue clusters are 277 birdies while red clusters are other candidates beyond birdies. Each blue point or red point represent a group of hits due to the big size of the waterfall plot.} Note that the background does not represent the raw data but data after Nebula, because the raw data is too big to plot at the same time. 
\label{fig:14}
\end{figure*}

With the clean data, we use DBSCAN to search for dense clusters and select clusters with the two rules above. The result is shown in Table \ref{tab:4}.  19 groups of birdies and 83 groups of candidates are selected. We are very happy to note that our pipeline found 277 birdies, 94.2177$\%$ of the total. Only one group of birdies is removed by Nebula. This means that we can successfully find candidate targets that match our desired characteristics. All of the candidates found are shown in Figure \ref{fig:14}. \par
\begin{figure*}[tbp]
\begin{minipage}[c]{1\linewidth}
\centering
\includegraphics[width = 1\linewidth]{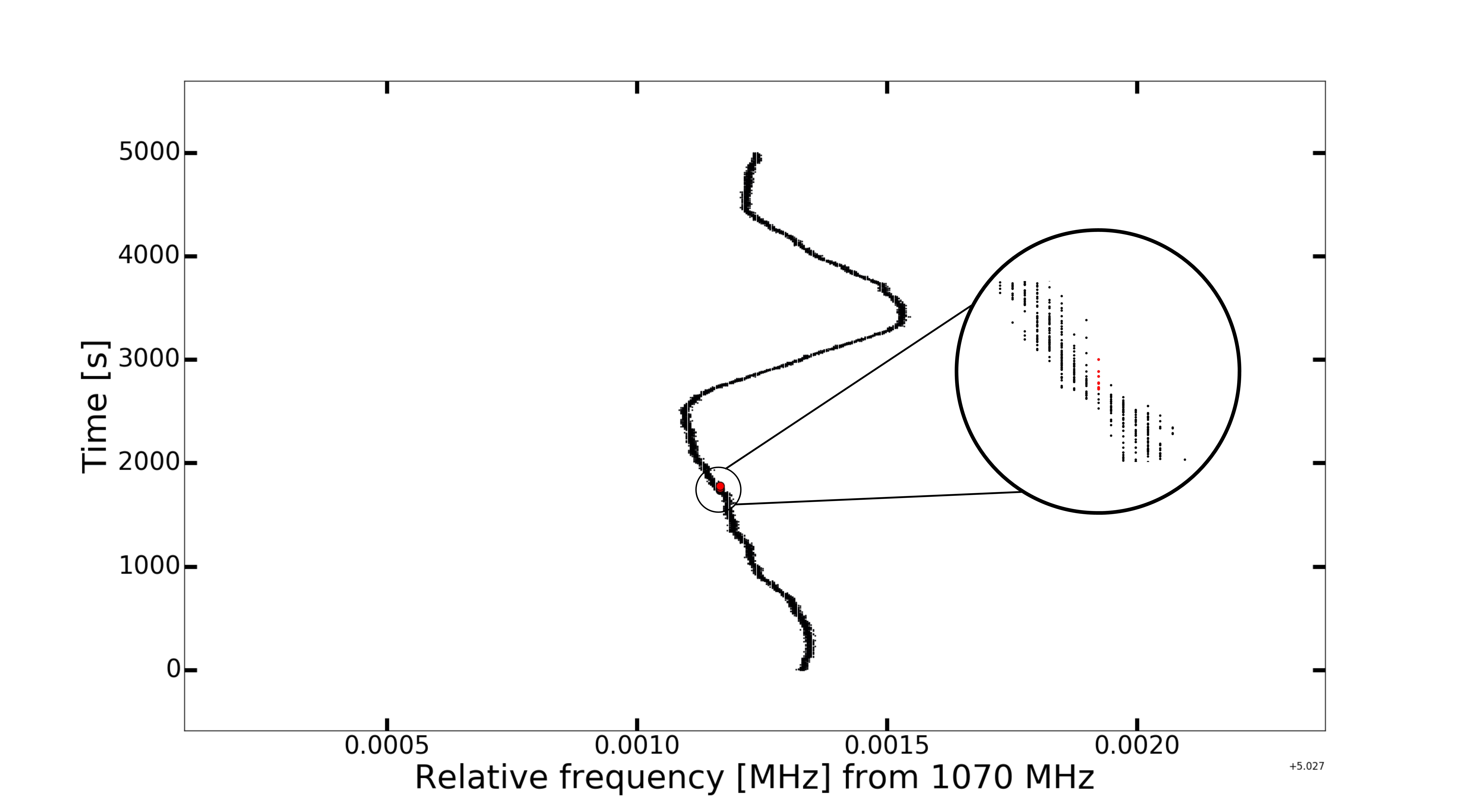}
\end{minipage}\\
\begin{minipage}[c]{1\linewidth}
\centering
\includegraphics[width = 1\linewidth]{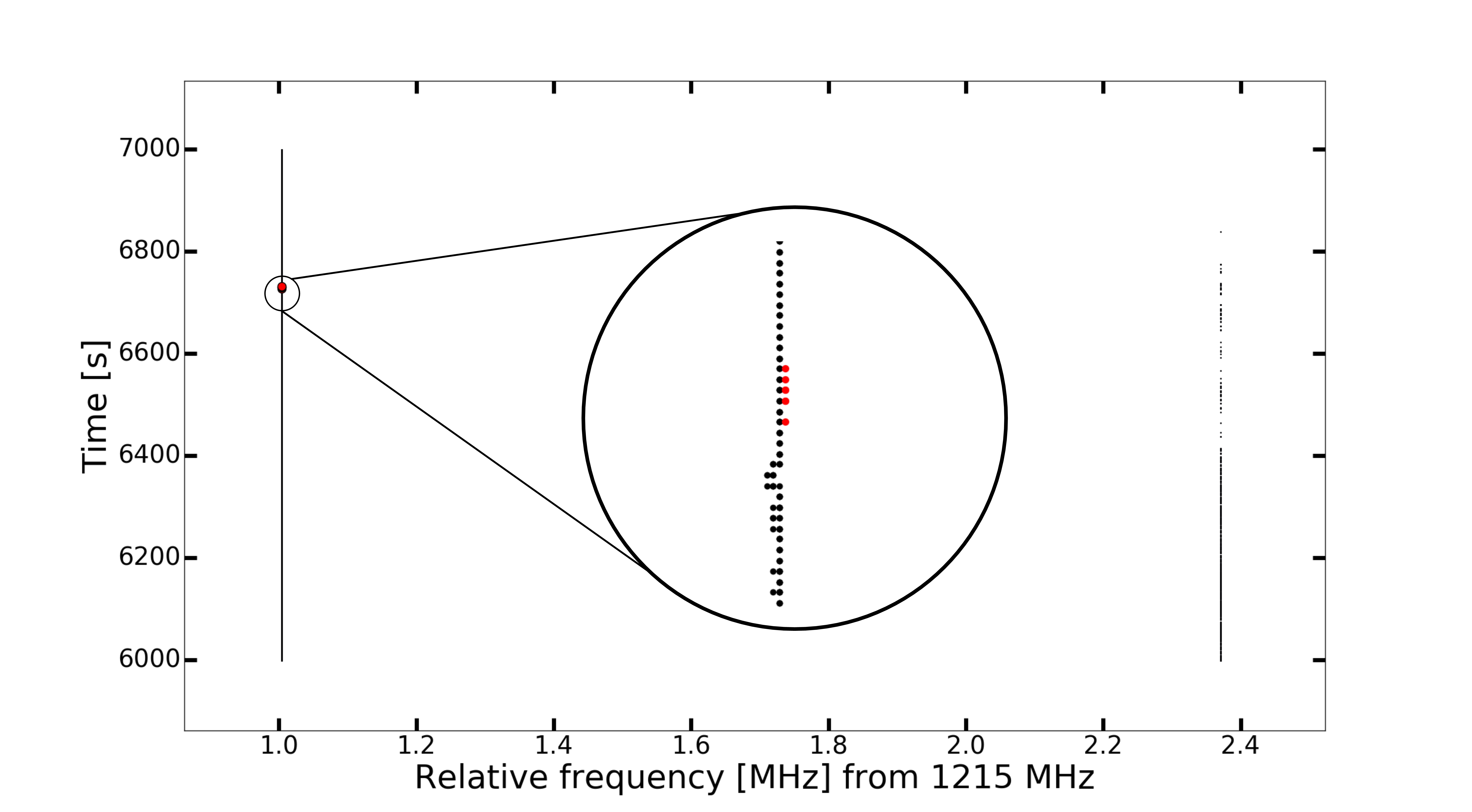}
\end{minipage}\\
\caption{Two groups of candidates within raw data. The red cluster in each figure is candidates while black points are raw data. It is clear that these candidates still belong to part of RFI. The one in top panel is drifting RFI while one in bottom panel is narrow-band RFI.}
\label{fig:15}
\end{figure*}
\begin{figure*}[tbp]
\begin{minipage}[c]{1\linewidth}
\centering
\includegraphics[width = 1\linewidth]{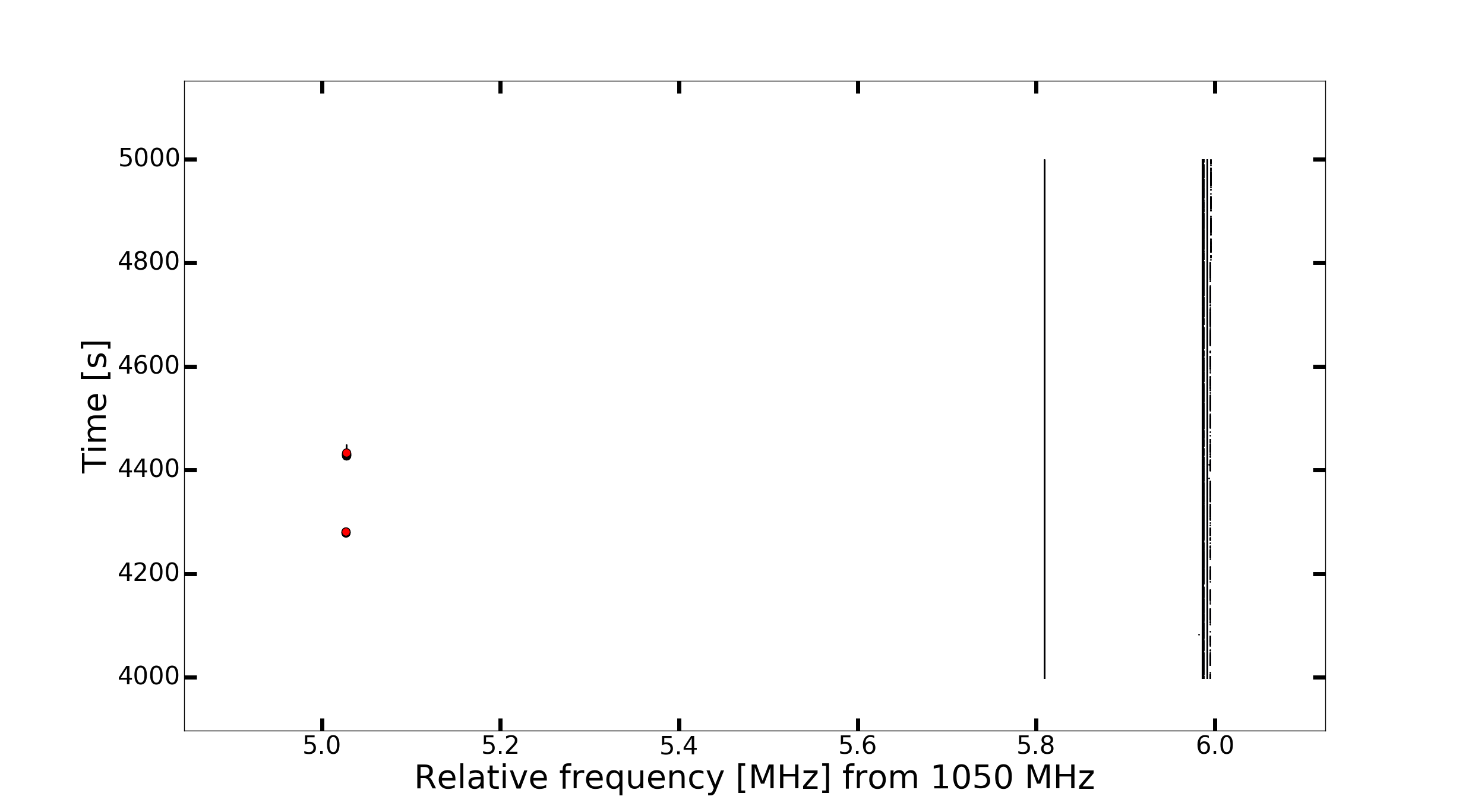}
\end{minipage}\\
\caption{Two groups of interesting candidates. The red group around 1055 MHz and 4280 seconds is called Group 1. The red group around 1055 MHz and 4430 seconds is called Group 2. From the figure, we can see that no other hits from nearby frequency channels in our five hour data.}
\label{fig:16}
\end{figure*}

In order to identify whether the candidates we found are unremoved RFI, we examine them with all hits found in the raw data to see if they are associated with RFI features. Four groups of candidates with raw data background are shown in Figure \ref{fig:15} and \ref{fig:16}. We examine all of the candidates and find that most of them have an obvious connection to RFI features. This is unsurprising because our RFI removal algorithm cannot completely remove all RFI. The unremoved RFI events match roughly to our definition of candidate's characteristics. Improvement of the RFI algorithm is an iterative process that we expect to continue as long as this system is used on FAST.\par
\begin{figure*}[tbp]
\begin{minipage}[c]{1\linewidth}
\centering
\includegraphics[width = 1\linewidth]{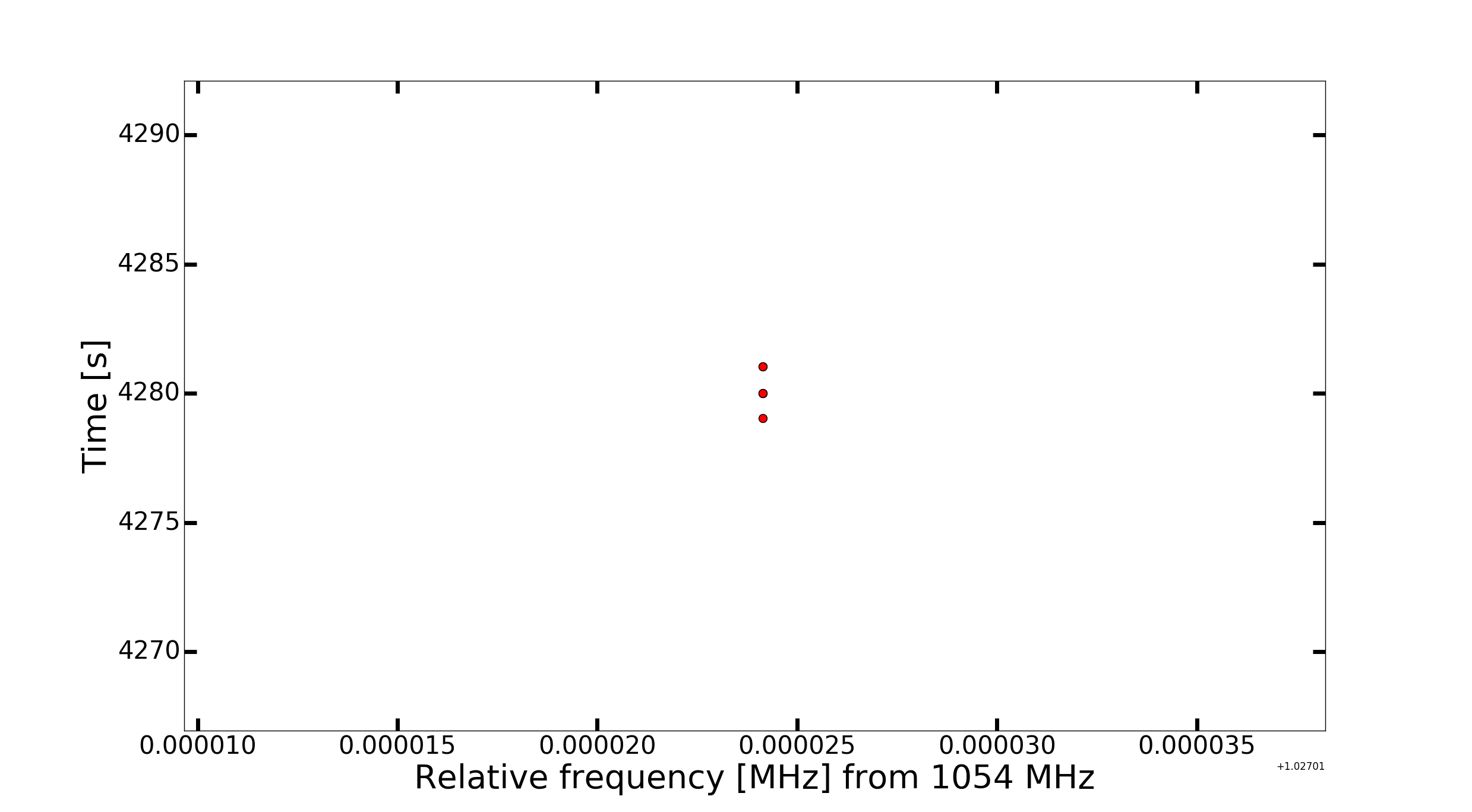}
\end{minipage}\\
\begin{minipage}[c]{1\linewidth}
\centering
\includegraphics[width = 1\linewidth]{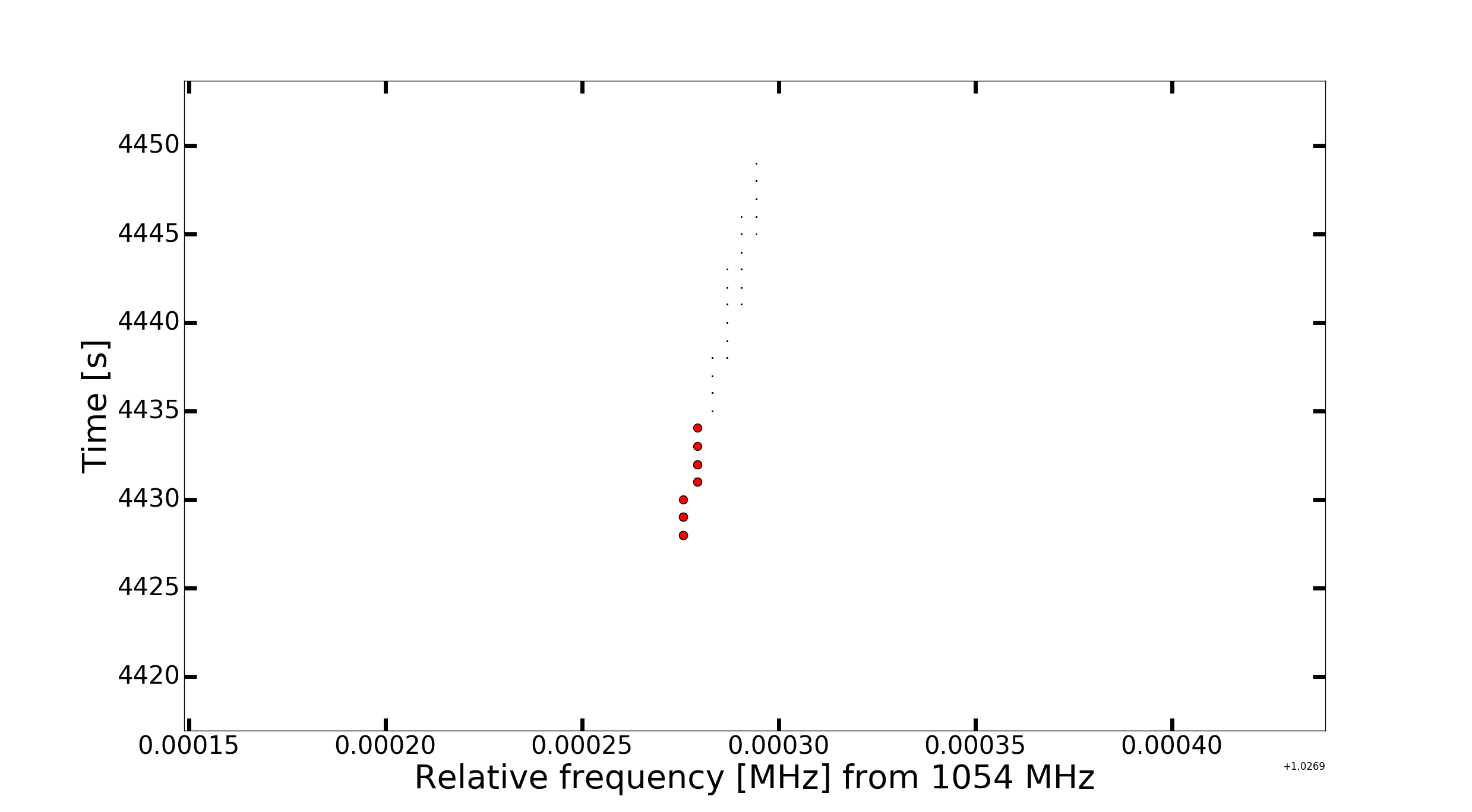}
\end{minipage}\\
\caption{Zoom in of candidates in Figure \ref{fig:16}. Group 1 in top panel only occupies one frequency channel. And Group 2 in bottom panel occupies six successive channels, totally $\sim$18.6 Hz of bandwidth. Note that Group 2 is in two colors, because only the red points are found by the SETI pipeline while black points are from the raw data.}
\label{fig:17}
\end{figure*}

 
For the five hours data, we find two group candidates of interest which have no hits around them. These two group candidates are shown in Figure \ref{fig:16}. The red group around 1055 MHz and 4280 seconds is called Group 1. The red group around 1055 MHz and 4430 seconds is called Group 2. We zoom into the two groups of candidates in Figure \ref{fig:17}. Group 1 is in the top panel of the figure with 5 hits in the same frequency channel. Group 2 is in the bottom panel of the figure with 80 hits in six successive channels. Group 1 is all from beam 15, and the duration is about 5 seconds. The duration of Group 2 is about 20 seconds. Events in beam 14 last 5 seconds, followed by 15 seconds of events in beam 15. \par
The two groups of candidates, even the other candidates which are verified be part of RFI, are very consistent with the two rules of ETI assumption. This result indicates that our data processing pipeline can pick out the signals that fit the assumption of ETI. Actually, this paper has presented the effectiveness of our method for RFI removal and candidate selection, which can certainly guarantee the successful implementation of observation for SETI with FAST.

\begin{table}[tbp]
    \setlength{\leftskip}{-20pt}
		\resizebox{0.5\textwidth}{!}{
			\begin{tabular}{lccc}
				\hline
                \hline
				Type of candidate & Birdies & Candidates & Total  \\
				\hline
				Number of hits &  277 & 593 & 870\\
				Number of groups & 19 & 83 & 102 \\
				\hline
			\end{tabular}
		}
		\\
		\caption{Result of candidate selection with DBSCAN. }
		\label{tab:4}
	\end{table}

%\begin{figure}[tbp]
%\begin{minipage}[c]{1\linewidth}
%\centering
%\includegraphics[width = 1\linewidth]{power.pdf}
%\end{minipage}\\
%\caption{Power of hits for group two. The left panel shows its waterfall, while the right plots detect power(top) and mean power(bottom), versus number of hits. {\bfseries There is no units for detect power and mean power, because all the power are calculated by the raw voltage data.}}
%\label{fig:18}
%\end{figure}

%\begin{figure}[tbp]
%\begin{minipage}[c]{1\linewidth}
%\centering
%\includegraphics[width = 1\linewidth]{gaussian_fitting_curve.pdf}
%\end{minipage}\\
%\caption{Gaussian fitting curve of the candidates in Figure \ref{fig:18}. We only focus on the hits in beam 15. They are more typical and fit the Gauss curve well.}
%\label{fig:19}
%\end{figure}

\subsection{The RFI environment of FAST}\label{sec:6.4}
As the largest single-aperture telescope in the world, the electromagnetic environment of FAST is a very important consideration. Due to its extreme sensitivity, it is a big challenge for FAST to mitigate RFI. The solutions for RFI mitigation of FAST include the Electromagnetic Compatibility measures of the telescope and the maintenance of radio-quiet zones around the site \citep{8738491}. Besides, a lot of RFI  monitoring for FAST has been done. Table \ref{tab:6} lists the known RFI sources. From the table, Civil aviation occupies 960-1215 MHz and ASIASTAR a geostationary satellite occupies 1467-1492 MHz. Some satellites occupy the middle frequency band, which only appear when they pass over or near the telescope. Unfortunately, the RFI monitoring doesn't have the same high-frequency resolution as our SETI back end. Some of the narrow-band RFI are not picked up by the monitor, also we can not mark the RFI with exact frequency channels now due to the low frequency resolution. It is an important and meaningful work to find the sources of these narrow-band RFI as the next step.

\begin{table*}[!htpb]
    \setlength{\leftskip}{-20pt}
		\resizebox{1\textwidth}{!}{
			\begin{tabular}{lcccccccc}
				\hline
                \hline
				Frequency band (MHz) & 960-1215 & 1176.45$\pm$1.023 & 1205-1209 & 1226.6-1229.6 & 1242-1250 & 1258-1278 & 1381-1386 & 1467-1492 \\
				\hline
				RFI Source &  The civil aviation & GPS L5, Galileo E5a & BD2 B2, Galileo E5b & GPS L2 & GLONASS L2 & BD2 B3 & GPS L3 & ASIASTAR \\
				\hline
			\end{tabular}
}
		\\
		\caption{RFI source list of FAST.}
		\label{tab:6}
	\end{table*}

\section{Conclusion and Future Plans}\label{sec:7}
FAST is the largest single-aperture telescope in the world; it's 19 beam receiver allows rapid and sensitive sky surveys with robust RFI rejection, ideal for SETI.\par
%FAST is the largest single-aperture telescope in the world, featuring high sensitivity and large observation area. Its unique hover feed cabin design is an innovation, greatly reducing the area of the telescope covered. The newly installed 19-beam receiver enables the telescope to observe more sky area at the same time. These technological innovations and advantages of FAST satisfy the requirements of SETI observation. \par
%The main conclusions we draw in the paper is below: \par
We have developed a SETI signal detection pipeline to operate on FAST's 19-beam seti instrument.  We conducted the first observational test of SETI with the 19-beam receiver of FAST in July, 2019.  By injecting test signals ("birdies") into our data, we demonstrated the pipeline's RFI removal capability. 
%The three kinds of RFI characterizing zone, broadband and drifting, are almost discarded by Nebula together Machine Learning. 
The ETI signal candidates were selected by the criteria outlined in Section \ref{sec:5}, and our data processing pipeline is working well on the preliminary data collected from FAST. Hopefully, we expect that these ETI signal candidates could come from some warm Earth-size planets in the Milky Way, the number of which can be roughly predicted by Drake equation\citep{1961PhT....14d..40D,2010LPICo1538.5211D,2018PASP..130e4101G}. 
\par

With the SETI capabilities demonstrated in this work, one can estimate the Equivalent Isotropic Radiated Power (EIRP)\citep{2017ApJ...849..104E} required of an alien transmitter to be detected by FAST. For a source of ETI signal candidates at at $d_{*}=$50 pc, a $\delta\nu=$ 4 Hz bandwidth, and a $t_{obs}=$ 0.25 second integration, the EIRP limit of FAST is $3.53\times 10^{13}$ W, given by:

\begin{equation}
EIRP = 4\pi d{_{*}}^{2}\sigma _{thresh}SEFD\sqrt{\frac{\delta \nu }{n_{pol}t_{obs}}}, 
\end{equation}
where the system equivalent flux density $SEFD = 2k_{B}T_{sys}/A_{eff}$ and $\sigma _{thresh}$=S/N$_{min}$ is a signal-to-noise threshold value.\par

We are planning to take data on FAST's commensal drift-scan surveys, such as the multiyear CRAFTS survey;
we also intend to observe targets from the Transiting Exoplanet Survey Satellite(TESS), as well as selected stars within $d_{*}=$50 pc from the sun, for example: \citep{Isaacson_2017}. \par

We are planning several improvements to the multi-year CRAFTS multi-beam SETI sky survey on FAST:\par
To improve FAST's sky survey sensitivity to narrow band signals we are working on upgrading the spectral resolution of the SERENDIP VI spectrometer to ~1 Hz channelization.\par
We also plan to continue improving our RFI mitigation and candidate selection post processing algorithms, which will benefit several other SETI sky surveys as well (SERENDIP6 sky surveys at Arecibo and Green Bank, and the SETI@home multi-beam sky survey at Arecibo).The main purpose of our program is to search for extraterrestrial civilizations,  but the project is also helpful to study the RFI environment around FAST.\par

We have begun working on real-time RFI rejection and first level identification of potential candidate signals, which would trigger a 100 second raw voltage dump of time domain data on all 38 signals from the the multibeam receiver (19 beams and two polarizations) for subsequent off-line analysis.   This time domain data would allow us to cross correlate beams, thereby reducing source position uncertainty, as well as provide more robust RFI mitigation.\par
We are also considering continuously recording raw time domain data streams from all 38 signal chains on the FAST multi-beam receiver, which will allow us to send out the data to SETI@home volunteers for a more thorough and sensitive analysis.  SETI@home is 20 times more sensitive to narrow band signals than SERENDIP,  because the enormous computing power provided by the SETI@home volunteers allows the SETI@home screensaver client to compute coherent very long duration spectra on tens of thousands of possible signal drift rates.  The SETI@home client also searches for pulses, signals with several different bandwidths, and signals that match the telescope beam patterns. An auto-correlation algorithm is used to search for repeating patterns.  Although SETI@home is more sensitive than SERENDIP, and searches for a very rich variety of signal types, we probably won't be able to record and process the full bandwidth of all the beams at FAST (recording 80 Gbits per second for several years is a lot to manage).  So SETI@home would likely process a part of the FAST multi-beam receiver band at very high sensitivity, while SERENDIP6 would process the full band at a reduced sensitivity.   
Longer term, FAST is planning a sensitive phased array feed,  which could provide roughly 100 simultaneous beams; excellent for a next generation SETI sky survey.\par
More generally, Earthlings are just beginning to learn how we might detect other civilizations if they are out there.   We've only had radio technology for a century; that's a blink of the eye in the history of the universe and life on this planet. We are beginning to explore tiny regions of the large parameter space of possible technosignatures from potential extraterrestrial civilizations\citep{2018AJ....156..260W}. Even though we are in an infant stage,  SETI science and technology is growing exponentially.  Radio telescope sensitivity has been doubling every 3.6 years for the last 60 years, and SETI spectometer capabilities have been doubling every 20 months for the last fourty years.  This SETI sky survey commissioning work is a significant step, leading to a powerful new SETI survey on FAST. 

\section{Acknowledgements}\label{sec:8}
We sincerely appreciate the referee’s rapid, thorough, and thoughtful response, which helped us greatly improve our manuscript. This work was supported by National Key R$\&$D Program of China (2017YFA0402600), the National Science Foundation of China (Grants No. 11573006, 11528306, 11803054, 11690024, 11725313), the China Academy of Sciences international Partnership Program No. 114A11KYSB20160008, Berkeley's Marilyn and Watson Alberts SETI Chair funds, the Berkeley SETI Research Center, and the Radio Astronomy Laboratory at Berkeley.  This work was also supported by National Astronomical Observatories, Chinese Academy of Sciences, and its FAST group. During this work EJK and JC were supported in part by donations from the Friends of SETI@home
(\href{http://setiathome.berkeley.edu}{setiathome.berkeley.edu}). Zhi-Song Zhang sincerely thank Ling-Jie Kong for her kindly helps on the response to the referee.

\software{DBSCAN \citep{Ester:1996:DAD:3001460.3001507}, scikit-learn \citep{scikit-learn}, Python 3 \citep{python, python2}, SciPy \citep{scipy},  NumPy \citep{numpy}, Astropy \citep{astropy1,astropy2},
Matplotlib \citep{2007CSE.....9...90H}, Nebula \citep{korpela19b} }

\bibliographystyle{aasjournal}
\bibliography{references}
\end{document}